\documentclass[a4paper,11pt]{article}
\pdfoutput=1 

\usepackage{jheppub} 

\makeatletter
\gdef\@fpheader{~\\[-30pt]}
\makeatother

\usepackage[T1]{fontenc} 

\usepackage{amssymb,amsmath}
\usepackage{slashed,bbold}
\usepackage{multirow}
\usepackage{enumitem}
\setenumerate[1]{label={\alph*)}, leftmargin=18pt}
\setitemize[1]{leftmargin=18pt}
\usepackage[table,dvipsnames]{xcolor}
\usepackage{hyperref}
\hypersetup{
	colorlinks = true,
	citecolor  = NavyBlue,
	linkcolor  = NavyBlue,
	urlcolor = NavyBlue
}
\usepackage{booktabs}
\usepackage{multirow}
\usepackage{fontawesome5}

\usepackage{tikz}
\usetikzlibrary{arrows.meta,positioning}
\usepackage{cleveref}
\crefname{table}{Table}{Tables}
\crefname{equation}{Eq.}{Eqs.}
\crefname{appendix}{App.}{Apps.}
\crefname{section}{Sec.}{Secs.}
\crefname{figure}{Fig.}{Figs.}

\newcommand{\s}{\hspace{0.8pt}}

\title{Logarithmic Wavelets for Dark Matter--Phonon Scattering}

\author{Xu-Xiang~Li}
\author{and Zhengkang~Zhang}
\affiliation{Department of Physics \& Astronomy, University of Utah, Salt Lake City, UT 84112, USA}

\emailAdd{xuxiang.li@utah.edu}
\emailAdd{z.k.zhang@utah.edu}

\abstract{
Phonon excitations in crystals are a promising detection channel for sub-GeV dark matter (DM), and anisotropic targets add directional sensitivity through the daily modulation of the rate. Exploiting these capabilities requires evaluating six-dimensional rate integrals across DM models, target materials, detector orientations, and times of day. The vector space integration method factorizes the calculation into projections of the DM velocity distribution and of the material response -- each computed once and reused -- contracted with an analytic kinematic matrix, reducing such scans to fast matrix algebra. In the phonon channel, however, the relevant momentum transfers span six orders of magnitude, and the linearly spaced Haar wavelet basis of existing implementations falls short: light mediator models demand an impractically large basis, and a single projection reused across DM masses loses its effective resolution for light DM. We introduce a logarithmic Haar wavelet basis that resolves both obstacles, and present a package \texttt{VectorPhonoDark}~\href{https://github.com/xuxiangli/VectorPhonoDark}{\faGithub} that implements the approach. On an Al$_2$O$_3$ daily modulation benchmark, it reproduces results from \texttt{PhonoDark}'s direct numerical integration while reducing the computational cost by orders of magnitude. Though developed here for phonons, the logarithmic wavelet basis generalizes to any DM detection channel spanning a wide range of momentum transfers, enabling efficient scans over DM models and detection strategies.
}

\begin{document}
\maketitle
\flushbottom
\setcounter{page}{2}

\newpage

\section{Introduction}
\label{sec:intro}

Sub-GeV dark matter (DM) is well motivated by a variety of theoretical scenarios, including freeze-in production~\cite{Hall:2009bx}, asymmetric DM~\cite{Kaplan:2009ag,Cohen:2010kn}, and strongly interacting massive particles~\cite{Hochberg:2014dra}, yet lies largely beyond the reach of conventional direct detection based on nuclear recoils; see Ref.~\cite{Zurek:2024qfm} for a recent review. A broad program has therefore emerged to search for light DM with low-threshold condensed matter systems~\cite{Kahn:2021ttr,Mitridate:2022tnv,Essig:2022dfa}. DM--electron scattering in semiconductors~\cite{Essig:2011nj,Essig:2015cda} probes DM masses down to $\sim 1$~MeV via eV-scale energy deposits~\cite{SENSEI:2024yyt,DAMIC-M:2025luv,SuperCDMS:2025dha}. For even lighter DM, electron excitation becomes kinematically inaccessible, and phonon excitations from the primary scattering process in crystal targets have been identified as a viable detection channel~\cite{Knapen:2017ekk,Griffin:2018bjn,Trickle:2019nya,Griffin:2019mvc}: with typical phonon energies of $\mathcal{O}(10)$~meV, this extends the reach down to DM masses of $\mathcal{O}(10)$~keV. Cryogenic calorimeters with the sub-eV energy resolution required for this channel are under active development~\cite{Anthony-Petersen:2024vdh,Angloher:2025fzw,Kennard:2026xlx}, and the first DM searches with eV- and sub-eV-scale energy thresholds have recently been reported~\cite{SuperCDMS:2020aus,CRESST:2024cpr,QROCODILE:2024nqm,TESSERACT:2025tfw}.

Beyond extending the mass reach, the phonon channel offers a qualitatively new handle on a putative signal: directionality. Many of the proposed target crystals are anisotropic, so the scattering rate depends on the orientation of the crystal axes relative to the DM wind, and modulates over a sidereal day as the Earth rotates~\cite{Griffin:2018bjn,Coskuner:2021qxo}.\footnote{Daily modulation from an anisotropic detector response has a longer history in other channels: early proposals exploited the direction-dependent scintillation light yield of organic crystals~\cite{Belli:1992zb,Bernabei:2003ct}; one of these, stilbene, has more recently been reconsidered for DM--electron scattering, where the anisotropy instead resides in the molecular excitation rate~\cite{Blanco:2021hlm}. Other proposals rely on ion channeling in crystalline targets~\cite{Avignone:2008cw,Creswick:2010dm,Bozorgnia:2011tk}, forward--backward discrimination of electrons ejected from two-dimensional targets such as graphene~\cite{Hochberg:2016ntt}, anisotropic electronic structure of Dirac materials~\cite{Coskuner:2019odd}, anisotropic Migdal matrix elements in oriented molecules~\cite{Blanco:2022pkt}, direction-dependent thresholds for defect production~\cite{Budnik:2017sbu,Kadribasic:2017obi}, and nuclear spin orientation in polarized targets~\cite{Chiang:2012ze,Catena:2018uae}.} Daily modulation tracking the phase of the DM wind is a powerful discriminant against backgrounds, which in low-threshold solid-state detectors are only partially characterized and in some cases remain unexplained~\cite{Baxter:2025odk}. A recent statistical analysis~\cite{Blanco:2026kda} found that the discovery significance of a modulating DM signal keeps growing with exposure even in the presence of an unknown (flat or modulating) background, and that optimizing the detector orientation can substantially reduce the required exposure. Exploiting this capability, however, is computationally demanding. The scattering rate is a six-dimensional integral over the incoming DM velocity $\boldsymbol{v}$ and the momentum transfer $\boldsymbol{q}$. A modest survey over, e.g., 3 benchmark DM models with 30 masses each, and 5 candidate materials requires 450 evaluations of the six-dimensional integral, and further scanning over $10^2$ detector orientations and 24 times of day to optimize the daily modulation sensitivity scales the number of evaluations up to $\sim 10^6$.

The vector space integration method, introduced in Refs.~\cite{Lillard:2023qlx,Lillard:2023cyy} and implemented in the \texttt{vsdm} package~\cite{Lillard:2025aim}, offers a way out of this scaling problem. The idea is to view the rate integral as an inner product in a function space, and turn the integration into a fast matrix multiplication between precomputed projection coefficients. Concretely, the integrand is a product of three factors: a $\boldsymbol{v}$-dependent DM velocity distribution function (VDF), a $\boldsymbol{q}$-dependent material form factor, and a DM mass-dependent factor that involves both $\boldsymbol{v}$ and $\boldsymbol{q}$ and enforces the kinematic constraints. The VDF and the material form factor are each projected onto a basis of wavelet-harmonic functions, and the remaining factor becomes a matrix which is diagonal in the angular indices and known analytically. The projections are independent of the DM mass by construction, and detector rotations act algebraically on the coefficients through Wigner matrices. The expensive numerical integrations are thus performed once per material per interaction type, and once per halo model, after which rates for arbitrary DM models, orientations, and times reduce to fast matrix multiplications, significantly accelerating orientation-intensive analyses. This design led Ref.~\cite{Lillard:2023qlx} to envision a community library of projection coefficients covering the detector materials and halo models of interest: the material-specific computation is done once, then shared and reused in any subsequent analysis.

Bringing this efficiency and reusability to the phonon channel is the goal of the present paper, and it turns out to require a new ingredient. The obstacle is a hierarchy of momentum scales peculiar to phonon excitations. With an energy threshold of $\omega_\text{min} \sim \mathcal{O}(\text{meV})$, the smallest momentum transfer that can contribute to the rate is $q_\text{min} = \omega_\text{min}/v_\text{max}$, of order an eV, where $v_\text{max} \sim 10^{-3}$ is the maximum DM speed in the lab frame. At the other end, the material form factor extends up to a UV cutoff enforced by the Debye--Waller factor, $q_\text{cut} \simeq 500$~keV, which is six orders of magnitude above $q_\text{min}$. Furthermore, for a given DM mass $m_\chi$, the kinematically allowed momentum transfers are bounded by $2 m_\chi v_\text{max}$, which can be much smaller than $q_\text{cut}$ for light DM. The Haar wavelet basis employed in Refs.~\cite{Lillard:2023qlx,Lillard:2023cyy,Lillard:2025aim}, whose supports are uniformly spaced in $q$, resolves structure at a single scale. This gives rise to two problems. First, for light mediator models, the integrand is highly peaked at low $q$, so capturing it with linearly spaced wavelets requires at least $q_\text{cut}/q_\text{min} \sim 10^6$ radial basis functions, far beyond what is computationally practical. Second, although a single projection extending to $q_\text{cut}$ formally serves all DM masses, only the fraction $k(m_\chi) = 2m_\chi v_\text{max}/q_\text{cut}$ of its basis functions covers the kinematically allowed region, so the effective resolution diminishes for light DM. Maintaining accuracy across the DM theory space then forces a prohibitively large basis and/or multiple projections tiling the mass range decade by decade, and we lose the efficiency and reusability that the method was designed for.

In this paper we show that both problems are solved by adopting Haar wavelets whose support boundaries are uniformly spaced in $\log q$ as opposed to $q$. Such a logarithmic wavelet basis devotes equal resolution to every decade of momentum transfer between $q_\text{min}$ and $q_\text{cut}$. For the convergence of the expansion, it reaches percent-level accuracy for light mediators with only $\mathcal{O}(100)$ radial wavelets while costing merely a factor of $\sim 4$ in basis size for heavy mediators. For the coverage of DM masses, the effective fraction of basis functions resolving the kinematically allowed region becomes $1 - \log k(m_\chi)/\log\epsilon$, with $\epsilon = q_\text{min}/q_\text{cut}$, instead of $k(m_\chi)$: the accuracy loss from reusing a single projection is logarithmic rather than linear in $k(m_\chi)$. As a result, one material form factor projection can cover the entire phonon-accessible mass range at sub-percent accuracy.

We implement the calculation in \texttt{VectorPhonoDark}~\href{https://github.com/xuxiangli/VectorPhonoDark}{\faGithub}~\cite{VectorPhonoDark}, the first complete pipeline applying the vector space integration method to DM--phonon scattering. The package interfaces with density functional theory (DFT) phonon data through \texttt{phonopy}~\cite{phonopy-phono3py-JPCM,phonopy-phono3py-JPSJ} and implements the projection of the VDF and of the material form factor (binned in energy deposit to accommodate the discrete phonon spectrum), on-the-fly evaluation of the analytic kinematic scattering matrix, and rate assembly for arbitrary detector orientations. Projections are stored as HDF5 files with full parameter metadata, following the same coefficient conventions as \texttt{vsdm}~\cite{Lillard:2025aim}; the files for GaAs and Al$_2$O$_3$, together with the scripts generating them from DFT inputs, are shipped with the package and provide the first phonon-channel entries of the coefficient library envisioned in Ref.~\cite{Lillard:2023qlx}. As a validation benchmark, we compute the daily modulation of the scattering rate in Al$_2$O$_3$ for the light dark photon mediator model, and find agreement with the direct numerical integration code \texttt{PhonoDark}~\cite{Trickle:2020oki,PhonoDark} at the $\lesssim 0.1\%$ level, while the full computation (4 masses $\times$ 24 time steps, fixed detector orientation) completes in $\sim 9$~seconds instead of $\sim 2.8$~hours on a single core of a personal computer -- a speedup of $\sim 1100$, after a one-time cost of $\sim 200$~seconds to compute the projection coefficients. Adding an orientation scan is essentially free in \texttt{VectorPhonoDark}, while it would multiply the cost of direct integration by the number of orientations, so the speedup grows with the size of the scan. Although we focus on phonons throughout, the logarithmic wavelet basis carries over to any detection channel involving a wide range of momentum transfers.

The rest of the paper is organized as follows. In \cref{sec:rate} we review the formalism of DM--phonon scattering, identify the three ingredients of the rate integral, and exhibit the momentum scale hierarchy described above. \Cref{sec:vector} reviews the vector space integration method. \Cref{sec:phonon} presents the main technical innovation of this work, the logarithmic wavelet basis, and shows how it overcomes the limitations of the linear basis when applied to the phonon channel. \Cref{sec:vectorphonodark} describes the design of \texttt{VectorPhonoDark} and the Al$_2$O$_3$ daily modulation benchmark. We conclude in \cref{sec:conclusion}, and present a detailed convergence study in \cref{app:convergence}. We focus on spin-independent interactions throughout this paper and leave the generalizations to other effective operators to future work.

\section{DM--phonon scattering rate}
\label{sec:rate}

Consider a DM particle with mass $m_\chi$ scattering off a crystal target composed of $N_T$ primitive cells. The differential scattering rate with respect to the energy deposit $\omega$ is given by~\cite{Trickle:2019nya,Lillard:2023cyy,Lillard:2025aim}:\footnote{Our normalization conventions follow Refs.~\cite{Lillard:2023cyy,Lillard:2025aim} and are related to those of Ref.~\cite{Trickle:2019nya} as follows: the total rate $R$ of \cref{eq:Rb_def} is that in a target of $N_T$ primitive cells, so dividing by the target mass $N_T\s m_\text{cell}$ recovers the rate per unit target mass of Ref.~\cite{Trickle:2019nya}; the material form factor is related to the dynamic structure factor $S(\boldsymbol{q},\omega)$ in Ref.~\cite{Trickle:2019nya} by $f_S^2 = \frac{\Omega}{2\pi} S$, where $\Omega$ is the volume of the primitive cell; and $\bar\sigma_0$ coincides with $\bar\sigma_n$ ($\bar\sigma_e$) defined there for DM coupling to nucleons (electrons).}
\begin{equation}
\label{eq:rate}
\frac{\mathrm{d}R}{\mathrm{d}\omega}
=
N_T \, \frac{\rho_\chi}{m_\chi} \, \frac{\bar\sigma_0}{4\pi \mu_\chi^2}
\int \mathrm{d}^3 \boldsymbol{v} \, \mathrm{d}^3 \boldsymbol{q} \;
g_\chi(\boldsymbol{v}, t) \,
F_\text{DM}^2(q) \,
f_S^2(\boldsymbol{q}, \omega) \,
\delta\biggl(\omega + \frac{q^2}{2 m_\chi} - \boldsymbol{q}\cdot\boldsymbol{v}\biggr) \,,
\end{equation}
where $\rho_\chi$ is the local DM energy density, $g_\chi(\boldsymbol{v}, t)$ is the DM VDF in the lab frame, $\bar\sigma_0$ is a reference cross section for DM scattering off a Standard Model (SM) particle, $F_\text{DM}(q)$ is the mediator form factor of the DM model, $\mu_\chi$ is the reduced mass of the DM and the SM particle it couples to (nucleon or electron depending on the model), and $f_S^2(\boldsymbol{q}, \omega)$ is the material form factor. The $\delta$-function enforces energy conservation for a non-relativistic DM particle: an incoming velocity $\boldsymbol{v}$ and a momentum transfer $\boldsymbol{q}$ deposit the energy $\omega = \boldsymbol{q}\cdot\boldsymbol{v} - q^2/(2 m_\chi)$ into the target. In practical calculations, we divide $\omega$ into bins of width $\Delta\omega$. The binned rate $R_b$ and the total rate $R$ are given by:
\begin{equation}
\label{eq:Rb_def}
R_b \equiv \int_{\omega_b - \Delta\omega/2}^{\omega_b + \Delta\omega/2} \mathrm{d}\omega \, \frac{\mathrm{d}R}{\mathrm{d}\omega} \,,
\qquad
R = \sum_b R_b \,,
\end{equation}
where $\omega_b$ is the bin center.

The integrand of \cref{eq:rate} factorizes into three independently specified inputs: the astrophysics of DM, encoded in $(\rho_\chi, g_\chi)$; the DM particle model, encoded in $(m_\chi, \mu_\chi, \bar\sigma_0, F_\text{DM})$; and the material response, encoded in $f_S^2$. The three inputs are coupled through the energy-conserving $\delta$-function. We now describe each ingredient in turn.

\paragraph{Astrophysics of DM.}
The astrophysical input to \cref{eq:rate} comprises the local DM energy density $\rho_\chi$ and the velocity distribution $g_\chi$. The energy density fixes the incident DM number density $\rho_\chi/m_\chi$, and hence the overall normalization of the rate; we adopt the commonly used value $\rho_\chi = 0.4~\text{GeV/cm}^3$. The lab-frame VDF is obtained from the galactic-frame VDF by a boost with the Earth's velocity $\boldsymbol{v}_\text{E}(t)$ in the galactic rest frame,
\begin{equation}
g_\chi(\boldsymbol{v}, t) = g_\chi^\text{gal}(\boldsymbol{v} + \boldsymbol{v}_\text{E}(t)) \,,
\end{equation}
where $g_\chi^\text{gal}$ is commonly assumed to be isotropic, i.e., to depend only on the DM speed in the galactic rest frame, and to vanish above the galactic escape velocity $v_\text{esc}$. The lab-frame DM speed is then bounded from above by
\begin{equation}
\label{eq:vmax}
v_\text{max} \equiv v_\text{esc} + v_\text{E} \sim 10^{-3} \, c \,.
\end{equation}
On the other hand, the energy-conserving $\delta$-function in \cref{eq:rate} fixes $\omega + q^2/(2 m_\chi) = \boldsymbol{q}\cdot\boldsymbol{v} \le qv$, so given $m_\chi$, $q$ and $\omega$, the DM speed is bounded from below by
\begin{equation}
\label{eq:vmin}
v_\text{min}(q, \omega) \equiv \frac{\omega}{q} + \frac{q}{2 m_\chi} \,.
\end{equation}

The benchmark choice of $g_\chi^\text{gal}$ is the standard halo model (SHM)~\cite{Drukier:1986tm}, a truncated Maxwell--Boltzmann distribution with velocity dispersion $v_0 = 230$~km/s, escape velocity $v_\text{esc} = 600$~km/s, and Earth speed $v_\text{E} = 240$~km/s, the same benchmark values used in Ref.~\cite{Trickle:2019nya}. Time variation of the direction of $\boldsymbol{v}_\text{E}$ relative to the crystal axes due to the Earth's rotation gives rise to daily modulation of the rate.

\paragraph{DM particle model.}
For a spin-independent interaction, it is conventional to characterize the overall strength of the DM--SM coupling by a reference cross section $\bar\sigma_0$. For $2\to 2$ scattering, $\bar\sigma_0$ is defined as the amplitude squared evaluated at a reference momentum transfer $q_\text{ref}$, multiplied by the phase space volume. The momentum dependence, which for tree-level processes originates solely from the mediator propagator, is captured by the mediator form factor $F_\text{DM}(q)$, normalized such that $F_\text{DM}(q_\text{ref}) = 1$:
\begin{equation}
\label{eq:FDM}
F_\text{DM}(q) =
\begin{cases}
1 & \text{(heavy mediator)}, \\[4pt]
(q_\text{ref}/q)^2 & \text{(light mediator)}.
\end{cases}
\end{equation}
The reference momentum is conventionally chosen as $q_\text{ref} = m_\chi v_0$ for DM coupling to nucleons, and $q_\text{ref} = \alpha m_e$ for DM coupling to electrons.

\paragraph{Material response.}
The material form factor is defined as
\begin{equation}
\label{eq:fS2}
f_S^2(\boldsymbol{q}, \omega) \equiv \frac{1}{N_T} \sum_f \bigl|\langle f| {\cal F}_T(\boldsymbol{q}) |i\rangle\bigr|^2 \, \delta(E_f - E_i - \omega) \,,
\end{equation}
where ${\cal F}_T(\boldsymbol{q})$ is the Fourier transform of the scattering potential normalized to the particle-level scattering amplitude introduced in Ref.~\cite{Trickle:2019nya}, and the sum runs over the target final states $|f\rangle$ with energies $E_f$, with $E_i$ the energy of the initial state $|i\rangle$. The $\delta$-function enforces energy conservation. For a discrete final state spectrum, integrating over an energy bin in \cref{eq:Rb_def} reduces to a sum over the final states whose excitation energies fall within that bin.

For single-phonon excitations in a crystal, the initial state $|i\rangle$ is the zero-phonon vacuum, and the final states are the one-phonon states $|\nu, \boldsymbol{k}\rangle$, labeled by branch $\nu = 1, \dots, 3n$ (for $n$ atoms per primitive cell) and momentum $\boldsymbol{k}$ in the first Brillouin zone (1BZ). Evaluating the matrix elements yields
\begin{equation}
\label{eq:fS2_phonon}
f_S^2(\boldsymbol{q}, \omega)
=
\sum_\nu \frac{1}{2\s\omega_{\nu, \boldsymbol{k}}}
\Biggl| \sum_j \frac{e^{-W_j(\boldsymbol{q})}}{\sqrt{m_j}} \, e^{i \boldsymbol{G} \cdot \boldsymbol{x}_j^0} \, \bigl(\boldsymbol{Y}_{\!j} \cdot \boldsymbol{\epsilon}_{\nu, \boldsymbol{k}, j}^*\bigr) \Biggr|^2
\delta(\omega - \omega_{\nu, \boldsymbol{k}}) \,,
\end{equation}
where $\boldsymbol{k}$ is the unique 1BZ momentum satisfying $\boldsymbol{q} = \boldsymbol{k} + \boldsymbol{G}$ with $\boldsymbol{G}$ a reciprocal lattice vector, $m_j$ and $\boldsymbol{x}_j^0$ are the mass and equilibrium position of the $j$-th atom in the primitive cell, $\omega_{\nu, \boldsymbol{k}}$ and $\boldsymbol{\epsilon}_{\nu, \boldsymbol{k}, j}$ are the phonon frequencies and polarization vectors, obtained from DFT calculations, and
\begin{equation}
W_j(\boldsymbol{q}) = \frac{1}{4 N_T m_j} \sum_{\nu, \boldsymbol{k}} \frac{|\boldsymbol{q}\cdot\boldsymbol{\epsilon}_{\nu, \boldsymbol{k}, j}|^2}{\omega_{\nu, \boldsymbol{k}}}
\label{eq:debye-waller}
\end{equation}
is the Debye--Waller factor. All the quantities above except the effective coupling vector $\boldsymbol{Y}_{\!j}$ are intrinsic properties of the target material; $\boldsymbol{Y}_{\!j}$ encodes how the DM couples to each atom $j$ in the primitive cell. For the two benchmark mediator models considered in this work,
\begin{equation}
\label{eq:Yj}
\boldsymbol{Y}_{\!j} =
\begin{cases}
\boldsymbol{q} \, A_j \, F_{N_j}(q) & \text{(hadrophilic scalar mediator)}, \\[8pt]
-\dfrac{\boldsymbol{q}\cdot\boldsymbol{Z}_j^\ast}{\hat{\boldsymbol{q}}\cdot\varepsilon_\infty\cdot\hat{\boldsymbol{q}}} & \text{(dark photon mediator)},
\end{cases}
\end{equation}
where $A_j$ is the mass number of the $j$-th atom, $F_{N_j}(q)$ is the nuclear form factor, $\boldsymbol{Z}_j^\ast$ is the Born effective charge tensor, $\hat{\boldsymbol{q}}$ is the unit vector in the direction of $\boldsymbol{q}$, and $\varepsilon_\infty$ is the high-frequency dielectric tensor.

At large momentum transfer, the Debye--Waller factor exponentially suppresses the single-phonon response, providing an effective UV cutoff in momentum. Viewing $\frac{1}{N_T}\sum_{\boldsymbol{k}}$ in \cref{eq:debye-waller} as an average over the 1BZ, we can estimate this momentum cutoff to be $\sim \sqrt{m_j \omega}$, where $\omega$ is a characteristic phonon frequency. Practically, we truncate the $q$ integration in \cref{eq:rate} at
\begin{equation}
\label{eq:qcut}
q_\text{cut} \equiv 10 \sqrt{m_\text{max}\, \omega_\text{max}} \,,
\end{equation}
where $m_\text{max}$ and $\omega_\text{max}$ are the largest atomic mass and phonon frequency, respectively, and the factor of $10$ is an arbitrary but conservative safety margin. Numerically, $q_\text{cut} \simeq 471$~keV for GaAs and $474$~keV for Al$_2$O$_3$, the two target materials studied in this work. For $m_\chi \lesssim 100$~MeV, the integrand vanishes before reaching this Debye--Waller cutoff due to the kinematic constraint $q < 2 m_\chi v_\text{max}$. We therefore define:
\begin{equation}
\label{eq:qmax}
q_\text{max} \equiv \min\bigl\{2 m_\chi v_\text{max} \,,\, q_\text{cut}\bigr\} \,.
\end{equation}
On the other hand, the momentum transfer needed to deposit an energy above a threshold $\omega_\text{min}$ is bounded from below by
\begin{equation}
\label{eq:qmin}
q_\text{min} \equiv \frac{\omega_\text{min}}{v_\text{max}}\,,
\end{equation}
as one can see by requiring $v_\text{min}(q, \omega_\text{min}) \leq v_\text{max}$. For $\omega_\text{min} = 1$~meV, $q_\text{min} \simeq 0.4$~eV, six orders of magnitude below the Debye--Waller cutoff. The wide range of momentum transfers relevant for DM--phonon scattering will turn out to be important for our discussion in later sections.

Directly evaluating the binned rates of \cref{eq:Rb_def} is straightforward but costly: each is a six-dimensional integral over $(\boldsymbol{v}, \boldsymbol{q})$ whose integrand requires DFT phonon data at every sampled $\boldsymbol{q}$ and a sum over the kinematically allowed final states. Worse, the entire computation must be repeated from scratch for every DM mass, mediator model, target material, detector orientation, and time of day in a parameter scan. The vector space integration method, to which we now turn, restructures the calculation so that each of the three ingredients is processed once and the results are recombined algebraically for any parameter combination, thus drastically reducing the computational cost.

\section{Scattering rate in the vector space integration method}
\label{sec:vector}

The scattering rate of \cref{eq:rate} is an integral of the product of two independently specified functions -- the VDF $g_\chi$ and the material form factor $f_S^2$ -- weighted by a kinematic kernel that carries the dependence on the DM particle model, up to an overall cross section normalization. The vector space integration method, introduced for DM direct detection calculations in Refs.~\cite{Lillard:2023qlx,Lillard:2023cyy} and implemented in the \texttt{vsdm} package~\cite{Lillard:2025aim}, takes advantage of this structure: the two functions are projected once onto a basis of the underlying function space, and the rate for any DM mass, mediator, detector orientation, and time is then assembled from the stored coefficients by fast matrix multiplications. In this section we review the method, following the conventions of Ref.~\cite{Lillard:2025aim}: \cref{sec:vsmethod} sets up the function space and the spherical Haar wavelet basis, and \cref{sec:ratevs} recasts the scattering rate as a matrix element in this basis. Applying the method to the phonon channel requires using a variant of the basis introduced in Refs.~\cite{Lillard:2023qlx,Lillard:2023cyy,Lillard:2025aim} due to the multi-scale nature of the problem, which we discuss in \cref{sec:phonon}.

\subsection{The vector space integration method}
\label{sec:vsmethod}

Consider an integral of the product of two real functions $f(x)$ and $g(x)$,
\begin{equation}
\langle f | g \rangle \equiv \int \mathrm{d}x \, f(x)\, g(x) \,,
\end{equation}
which can be viewed as an inner product of two vectors in the space of square-integrable functions, as reflected by the bra-ket notation. We can choose a complete orthonormal basis of this space: a set of real functions $\phi_n(x)$ with $n = 0, 1, \dots$ (denoted as $|n\rangle$ in the bra-ket form) satisfying the orthonormality and completeness relations
\begin{equation}
\label{eq:basis1d}
\int \mathrm{d}x \, \phi_n(x) \phi_{n'}(x) = \delta_{nn'} \,, \qquad \sum_n \phi_n(x) \phi_n(x') = \delta(x - x') \,.
\end{equation}
The completeness relation allows any function in the space to be expanded in this basis,
\begin{equation}
\label{eq:expansion1d}
f(x) = \sum_n \langle n | f \rangle \, \phi_n(x) \,,
\qquad
\text{with}
\quad
\langle n | f \rangle = \int  \mathrm{d}x \, \phi_n(x) \, f(x) \,.
\end{equation}
Note that since both $\phi_n$ and $f$ are real functions, $\langle n | f \rangle = \langle f | n \rangle$. Expanding both $f$ and $g$ in $\langle f | g \rangle$ and using the orthonormality relation then rewrites the integral as a sum of products of coefficients,
\begin{equation}
\label{eq:parseval}
\langle f | g \rangle = \sum_n \langle f | n \rangle \langle n | g \rangle \,.
\end{equation}
If the basis is well adapted to the functions at hand, the sum converges after a moderate number of terms, and the coefficients $\langle n | f \rangle$ and $\langle n | g \rangle$, once computed, can be reused in every integral involving $f$ or $g$.

For the scattering rate we need the three-dimensional version of this construction, applied to the velocity and momentum integrations. Following Ref.~\cite{Lillard:2025aim}, for functions of a three-dimensional variable $\boldsymbol{u}$ supported on the ball $|\boldsymbol{u}| \leq u_\text{max}$ we define the inner product with a normalized measure,
\begin{equation}
\label{eq:inner3d}
\langle f | g \rangle \equiv \int \frac{\mathrm{d}^3\boldsymbol{u}}{u_\text{max}^3} \, f(\boldsymbol{u})\, g(\boldsymbol{u}) \,.
\end{equation}
A complete orthonormal basis of this space is a set of real functions $\phi_{n\ell m}(\boldsymbol{u})$, labeled by a radial index $n = 0, 1, \dots$ and angular indices $\ell, m$ (denoted as $|n\ell m\rangle$), satisfying the orthonormality and completeness relations analogous to \cref{eq:basis1d},
\begin{equation}
\label{eq:basis3d_relations}
\int \frac{\mathrm{d}^3\boldsymbol{u}}{u_\text{max}^3} \, \phi_{n\ell m}(\boldsymbol{u})\, \phi_{n'\ell'm'}(\boldsymbol{u}) = \delta_{nn'}\, \delta_{\ell\ell'}\, \delta_{mm'} \,,
\quad
\sum_{n\ell m} \phi_{n\ell m}(\boldsymbol{u})\, \phi_{n\ell m}(\boldsymbol{u}') = u_\text{max}^3\, \delta^3(\boldsymbol{u} - \boldsymbol{u}') \,.
\end{equation}
Any function $f(\boldsymbol{u})$ on the ball is then expanded in this basis,
\begin{equation}
\label{eq:expansion3d}
f(\boldsymbol{u}) = \sum_{n\ell m} \langle n \ell m | f \rangle \, \phi_{n\ell m}(\boldsymbol{u}) \,,
\qquad
\text{with}
\quad
\langle n \ell m | f \rangle = \int \frac{\mathrm{d}^3\boldsymbol{u}}{u_\text{max}^3} \, \phi_{n\ell m}(\boldsymbol{u})\, f(\boldsymbol{u}) \,,
\end{equation}
and the inner product of two functions reduces to a sum of products of coefficients,
\begin{equation}
\label{eq:parseval3d}
\langle f | g \rangle = \sum_{n\ell m} \langle f | n \ell m \rangle \langle n \ell m | g \rangle \,,
\end{equation}
exactly as in the one-dimensional case of \cref{eq:expansion1d,eq:parseval}. With the normalized measure of \cref{eq:inner3d}, the basis functions $\phi_{n\ell m}$ are dimensionless, and the coefficients $\langle n\ell m | f \rangle$ carry the units of $f$ itself.

Concretely, we consider basis functions that factorize into radial and angular parts,
\begin{equation}
\label{eq:basis3d}
\phi_{n\ell m}(\boldsymbol{u}) = h_n(u/u_\text{max})\, Y_{\ell m}(\hat{\boldsymbol{u}}) \,,
\end{equation}
where the $Y_{\ell m}$ are real spherical harmonics, related to the standard complex ones by
\begin{equation}
Y_{\ell m}(\theta, \phi) =
\begin{cases}
\sqrt{2}\, (-1)^m \operatorname{Im} Y_{\ell}^{|m|}(\theta, \phi) & m < 0 \,, \\
Y_{\ell}^{0}(\theta, \phi) & m = 0 \,, \\
\sqrt{2}\, (-1)^m \operatorname{Re} Y_{\ell}^{m}(\theta, \phi) & m > 0 \,,
\end{cases}
\end{equation}
and the $h_n$ are real radial functions, orthonormal on the unit interval with respect to the radial measure,
\begin{equation}
\label{eq:radial_orthonormality}
\int_0^1 x^2\, \mathrm{d}x \, h_n(x)\, h_{n'}(x) = \delta_{nn'} \,.
\end{equation}

The rotational symmetry of the problem (that the rate is invariant under simultaneous rotations of the DM wind and the target) makes the real spherical harmonics the preferred angular basis, while we are free to choose the radial functions $h_n$, subject to the orthonormality condition of \cref{eq:radial_orthonormality}. The choice made in Refs.~\cite{Lillard:2023qlx,Lillard:2023cyy,Lillard:2025aim} is a spherical Haar wavelet basis. For $n \geq 1$ the wavelet $h_n(x)$ is a pair of concatenated top-hat functions,
\begin{equation}
\label{eq:haar_def}
h_n(x) = \begin{cases}
+ A_n & x_{1,n} \leq x < x_{2,n} \,, \\
- B_n & x_{2,n} \leq x < x_{3,n} \,, \\
0 & \text{otherwise} \,,
\end{cases}
\end{equation}
while $h_0 = A_0$ is constant on the domain of the basis. For $n\ge 1$, the index $n = 2^\lambda + \mu$ is organized by a generation label $\lambda = 0, 1, \dots$ and a position label $\mu = 0, 1, \dots, 2^\lambda - 1$; each generation subdivides the supports of the previous one, with the wavelets $(\lambda+1, 2\mu)$ and $(\lambda+1, 2\mu+1)$ occupying the lower and upper parts of the support of $(\lambda, \mu)$. The normalizations $A_n, B_n > 0$ are fixed by \cref{eq:radial_orthonormality} in terms of the support boundaries $x_{1,n} < x_{2,n} < x_{3,n}$, which are therefore the only remaining freedom. Refs.~\cite{Lillard:2023qlx,Lillard:2023cyy,Lillard:2025aim} space them uniformly on $[0, 1]$,
\begin{equation}
\label{eq:lin_bounds}
x_{1,n} = 2^{-\lambda}\mu \,, \qquad x_{2,n} = 2^{-\lambda}\Bigl(\mu + \frac{1}{2}\Bigr) \,, \qquad x_{3,n} = 2^{-\lambda}(\mu + 1) \,,
\end{equation}
which we refer to as the \emph{linear} wavelet basis. Orthonormality then fixes $A_0 = \sqrt{3}$ for $h_0$ on the domain $[0, 1]$, while for $n \geq 1$,
\begin{equation}
\label{eq:haar_norm}
A_{n} = \sqrt{\frac{3}{x_{3,n}^3 - x_{1,n}^3} \, \frac{x_{3,n}^3 - x_{2,n}^3}{x_{2,n}^3 - x_{1,n}^3}} \,,
\qquad B_{n} = \frac{x_{2,n}^3 - x_{1,n}^3}{x_{3,n}^3 - x_{2,n}^3} \, A_n \,.
\end{equation}
The first eight basis functions are illustrated in the left panel of \cref{fig:wavelet_shapes} below.

\subsection{The scattering rate as a matrix element}
\label{sec:ratevs}

We now apply the vector space integration method to the scattering rate of \cref{eq:rate}: the VDF and the material form factor are each expanded in the basis of \cref{eq:basis3d}, and the kinematic kernel between them becomes a matrix. Concretely, defining:
\begin{equation}
\label{eq:kernel_def}
{\cal O}_\omega(\boldsymbol{v}, \boldsymbol{q}) \equiv \frac{F_\text{DM}^2(q)}{4\pi \mu_\chi^2 m_\chi}\, \delta\Bigl(\omega + \frac{q^2}{2m_\chi} - \boldsymbol{q}\cdot\boldsymbol{v}\Bigr) \,,
\end{equation}
we can rewrite the rate formula as:
\begin{align}
\frac{\mathrm{d}R}{\mathrm{d}\omega} &=
N_T\, \rho_\chi\, \bar\sigma_0\, v_\text{max}^3q_\text{max}^3
\int \frac{\mathrm{d}^3\boldsymbol{v}}{v_\text{max}^3} \frac{\mathrm{d}^3\boldsymbol{q}}{q_\text{max}^3} \,
g_\chi(\boldsymbol{v}) \,
{\cal O}_\omega(\boldsymbol{v}, \boldsymbol{q})\,
f_S^2(\boldsymbol{q}, \omega)
 \nonumber\\[5pt]
&= N_T\, \rho_\chi\, \bar\sigma_0\, v_\text{max}^3q_\text{max}^3
\sum_{n\ell m} \sum_{n'\ell'm'}
\langle g_\chi | n \ell m \rangle \,
\langle n\ell m| {\cal O}_\omega | n' \ell' m' \rangle \,
\langle n' \ell' m' | f_S^2(\omega) \rangle \,,
\end{align}
where
\begin{equation}
\langle n \ell m |\, {\cal O}_\omega \,| n' \ell' m' \rangle
\equiv
\int \frac{\mathrm{d}^3\boldsymbol{v}}{v_\text{max}^3} \frac{\mathrm{d}^3\boldsymbol{q}}{q_\text{max}^3} \,
\phi_{n\ell m}(\boldsymbol{v})\, {\cal O}_\omega(\boldsymbol{v}, \boldsymbol{q})\, \phi_{n'\ell'm'}(\boldsymbol{q})\,.
\end{equation}
Using $\int\mathrm{d}\Omega_v \, \mathrm{d}\Omega_q \, Y_{\ell m}(\hat{\boldsymbol{v}})\,\delta(\cos\theta_{vq}-z)\,Y_{\ell' m'}(\hat{\boldsymbol{q}}) = 2\pi\,P_{\ell}(z)\,\delta_{\ell\ell'}\delta_{mm'}$ to carry out the angular integrals, we obtain:
\begin{equation}
\label{eq:rate_nlm_diff}
\frac{\mathrm{d}R}{\mathrm{d}\omega}
=
N_T\, \rho_\chi\, \bar\sigma_0\, \frac{v_\text{max}^5}{q_\text{max}}
\sum_{\ell m} \sum_{n n'}
\langle g_\chi | n \ell m \rangle \,
{\cal I}^{(\ell)}_{nn'}(\omega) \,
\langle n' \ell m | f_S^2(\omega) \rangle \,,
\end{equation}
where
\begin{align}
\label{eq:Ilnn}
{\cal I}^{(\ell)}_{nn'}(\omega)
&=
\frac{q_\text{max}^3/v_\text{max}^3}{2 m_\chi \mu_\chi^2}
\int_{q_\text{min}}^{q_\text{max}} \frac{q\, \mathrm{d}q}{q_\text{max}^2} \,
h_{n'}(q/q_\text{max})\,
F_\text{DM}^2(q) \nonumber \\
&\hspace{60pt}\times
\int_{v_\text{min}(q,\, \omega)}^{v_\text{max}} \frac{v\, \mathrm{d}v}{v_\text{max}^2} \,
h_{n}(v/v_\text{max}) \,
P_{\ell}\biggl(\frac{v_\text{min}(q, \omega)}{v}\biggr)
\end{align}
is dimensionless and called the \emph{kinematic scattering matrix} in Refs.~\cite{Lillard:2023cyy,Lillard:2025aim}. The limits of integration are set by the kinematic and Debye--Waller constraints discussed in \cref{sec:rate}; see \cref{eq:vmax,eq:vmin,eq:qmax,eq:qmin}. Crucially, since ${\cal O}_\omega (\boldsymbol{v}, \boldsymbol{q})$ is invariant under rotations -- it depends only on $v$, $q$ and $\hat{\boldsymbol{q}}\cdot\hat{\boldsymbol{v}}$ -- its matrix elements are diagonal in the angular indices and independent of $m$, leaving the reduced matrix ${\cal I}^{(\ell)}_{nn'}$. For the piecewise-constant Haar wavelets $h_n$ introduced in \cref{sec:vsmethod}, the double integral in \cref{eq:Ilnn} has a closed-form expression, given in Appendix~B of Ref.~\cite{Lillard:2023cyy}. The kinematic matrix is therefore analytic and inexpensive: no numerical integration and no material data are involved in its evaluation.

Integrating \cref{eq:rate_nlm_diff} over an energy bin gives the binned rate of \cref{eq:Rb_def}. Because $f_S^2(\boldsymbol{q}, \omega)$ is a sum of $\delta$-functions at the mode energies, this integration over $\omega$ is in fact a sum over final states. This sum couples the material form factor back to the kinematic matrix ${\cal I}^{(\ell)}_{nn'}(\omega)$ through their common dependence on $\omega$, spoiling the clean factorization at the differential rate level. We therefore approximate ${\cal I}^{(\ell)}$ as constant across each bin, ${\cal I}^{(\ell)}(\omega) \simeq {\cal I}^{(\ell)}(\omega_b)$, where $\omega_b$ is the bin center. With ${\cal I}^{(\ell)}(\omega_b)$ pulled out of the energy integral, the binned rate takes the form
\begin{equation}
\label{eq:rate_nlm}
R_b
=
\int_{\omega_b - \Delta\omega/2}^{\omega_b + \Delta\omega/2} \mathrm{d}\omega \, \frac{\mathrm{d}R}{\mathrm{d}\omega}
\simeq
N_T\, \rho_\chi\, \bar\sigma_0\, \frac{v_\text{max}^5}{q_\text{max}}
\sum_{\ell m} \sum_{n n'}
\langle g_\chi | n \ell m \rangle \,
{\cal I}^{(\ell)}_{nn'}(\omega_b) \,
\langle n' \ell m | f_{S,b}^2 \rangle \,,
\end{equation}
where
\begin{equation}
\label{eq:ff_binned}
f_{S,b}^2(\boldsymbol{q}) \equiv \int_{\omega_b - \Delta\omega/2}^{\omega_b + \Delta\omega/2} \mathrm{d}\omega \, f_S^2(\boldsymbol{q}, \omega)
\end{equation}
is the \emph{binned material form factor}, a finite sum over the phonon modes with $\omega_{\nu,\boldsymbol{k}}$ inside bin $b$. The factorization is now complete: the rate involves one kinematic scattering matrix per energy bin, and all the material data collapse into the coefficients $\langle n \ell m | f_{S,b}^2 \rangle$, which can be computed once and for all.

Detector orientations and times of day enter through rotations, which act simply on the expansion coefficients. A rotation ${\cal R}$ mixes the real spherical harmonics within each multiplet,
\begin{equation}
\label{eq:G_def}
{\cal R} \cdot Y_{\ell m}(\hat{\boldsymbol{u}}) \equiv Y_{\ell m}({\cal R}^{-1}\hat{\boldsymbol{u}}) = \sum_{m'} G^{(\ell)}_{m'm}({\cal R})\, Y_{\ell m'}(\hat{\boldsymbol{u}}) \,,
\end{equation}
where $G^{(\ell)}_{m'm}({\cal R})$ is the Wigner $G$-matrix, the real-basis analogue of the Wigner $D$-matrix, whose explicit expression in terms of $D^{(\ell)}_{m'm}$ is given in Appendix A of Ref.~\cite{Lillard:2023cyy}. A change of detector orientation is a rotation of the crystal, i.e., of the material form factor, $f_{S,b}^2(\boldsymbol{q}) \to {\cal R} \cdot f_{S,b}^2(\boldsymbol{q})$, equivalent to transforming its coefficients as
\begin{equation}
\label{eq:rot_ff}
\langle n \ell m | {\cal R} \cdot f_{S,b}^2 \rangle = \sum_{m'} G_{mm'}^{(\ell)}({\cal R}) \, \langle n \ell m' | f_{S,b}^2 \rangle \,.
\end{equation}
Daily modulation is instead a rotation of the VDF. Ignoring the Earth's orbital motion within a day, $\boldsymbol{v}_\text{E}(t) = {\cal R}(t)\, \boldsymbol{v}_\text{E}(0)$, with ${\cal R}(t)$ capturing the Earth's rotation. Isotropy of the galactic-frame VDF then gives
\begin{equation}
\label{eq:rot_vdf}
g_\chi(\boldsymbol{v}, t) = g_\chi^\text{gal}\bigl(\boldsymbol{v} + {\cal R}(t)\, \boldsymbol{v}_\text{E}(0)\bigr) = g_\chi\bigl({\cal R}^{-1}(t)\, \boldsymbol{v}, \, 0\bigr) \,,
\end{equation}
so the time dependence of the rate is obtained by rotating the VDF coefficients computed once at $t=0$. Collecting everything, the binned rate for an arbitrary relative orientation ${\cal R}$ (applied to either ingredient) is
\begin{equation}
\label{eq:rate_master}
R_b({\cal R})
=
N_T\, \rho_\chi\, \bar\sigma_0\, \frac{v_\text{max}^2}{q_\text{max}}
\sum_{\ell}\sum_{m m'} G_{mm'}^{(\ell)}({\cal R}) \; {\cal K}_{mm'}^{(\ell)}(\omega_b) \,,
\end{equation}
with the \emph{partial rate matrix}~\cite{Lillard:2023qlx,Lillard:2023cyy,Lillard:2025aim}
\begin{equation}
\label{eq:K_def}
{\cal K}_{mm'}^{(\ell)}(\omega_b) = v_\text{max}^3 \sum_{nn'}
\langle g_\chi | n \ell m \rangle \,
{\cal I}_{nn'}^{(\ell)}(\omega_b) \,
\langle n' \ell m' | f_{S,b}^2 \rangle \,.
\end{equation}

\Cref{eq:rate_master,eq:K_def} exhibit the division of labor that makes the method efficient for scans. The expensive objects are the projection coefficients: computing $\langle n\ell m | f_{S,b}^2 \rangle$ requires three-dimensional numerical integrals over DFT phonon data for every energy bin. These coefficients can be computed \emph{once per material and coupling type}, regardless of $m_\chi$ or the mediator mass, just as $\langle n\ell m | g_\chi \rangle$ is computed once per halo model. The kinematic scattering matrix ${\cal I}_{nn'}^{(\ell)}(\omega_b)$ must be evaluated for each DM mass and mediator, but it is analytic and adds a negligible cost. Detector orientations and times require no new integrals at all: each amounts to contracting the precomputed partial rate matrix with a new $G^{(\ell)}({\cal R})$, at a cost of microseconds per configuration. A scan over $N_m$ masses, $N_\text{med}$ mediators, $N_\text{mat}$ materials, $N_{\cal R}$ orientations, and $N_t$ times therefore requires only one set of material form factor projections -- one per energy bin -- per material and coupling type, instead of $N_m N_\text{med} N_\text{mat} N_{\cal R} N_t$ six-dimensional integrals. For orientation-intensive analyses the overall speedup can reach $\sim 10^8$~\cite{Lillard:2023cyy}.

For this strategy to deliver, however, the basis expansion must converge with a manageable number of coefficients $\langle n\ell m | f_{S,b}^2 \rangle$ for \emph{all} the DM models in the scan, and the stored projections must retain their accuracy for all the masses. Both requirements depend on the choice of radial basis functions $h_n$. The linear wavelets work well for DM--electron scattering where the relevant momenta span a relatively narrow range, but they fail in the phonon channel, which covers a much wider range of momenta; we address this issue in the next section.

\section{Logarithmic wavelets for the phonon channel}
\label{sec:phonon}

As established in \cref{sec:rate}, the momentum transfers relevant for DM--phonon scattering can span up to six orders of magnitude: from $q_\text{min} \simeq 0.4$~eV, set by meV-scale thresholds, up to the Debye--Waller cutoff $q_\text{cut} \simeq 500$~keV. Over such a wide range, the linear basis faces two obstacles: (i) for light mediator models, the integrand is highly peaked toward low $q$, so it requires an impractically large number of basis functions to resolve the dominant contribution near $q_\text{min}$; (ii) if we want to reuse a single material form factor projection across the full DM mass range, we have to set $q_\text{max}$ in \cref{eq:qmax} to the Debye--Waller cutoff $q_\text{cut}$, which means that for $m_\chi \ll 100$~MeV, the integrand only has support over a small fraction of the total range $\bigl[q_\text{min}, q_\text{max}\bigr]$, and the effective resolution is degraded. In this section, we overcome both obstacles by replacing the linear Haar wavelet basis with a logarithmic Haar wavelet basis, which devotes equal resolution to every decade of momentum transfer. We construct the logarithmic wavelet basis in \cref{sec:log}, and then show in \cref{sec:conv,sec:universal} that it resolves both obstacles -- fast convergence for all mediator models, and a single projection that serves all DM masses.

\subsection{Logarithmic wavelets}
\label{sec:log}

\begin{figure}[t]
\centering
\includegraphics[width=0.95\textwidth]{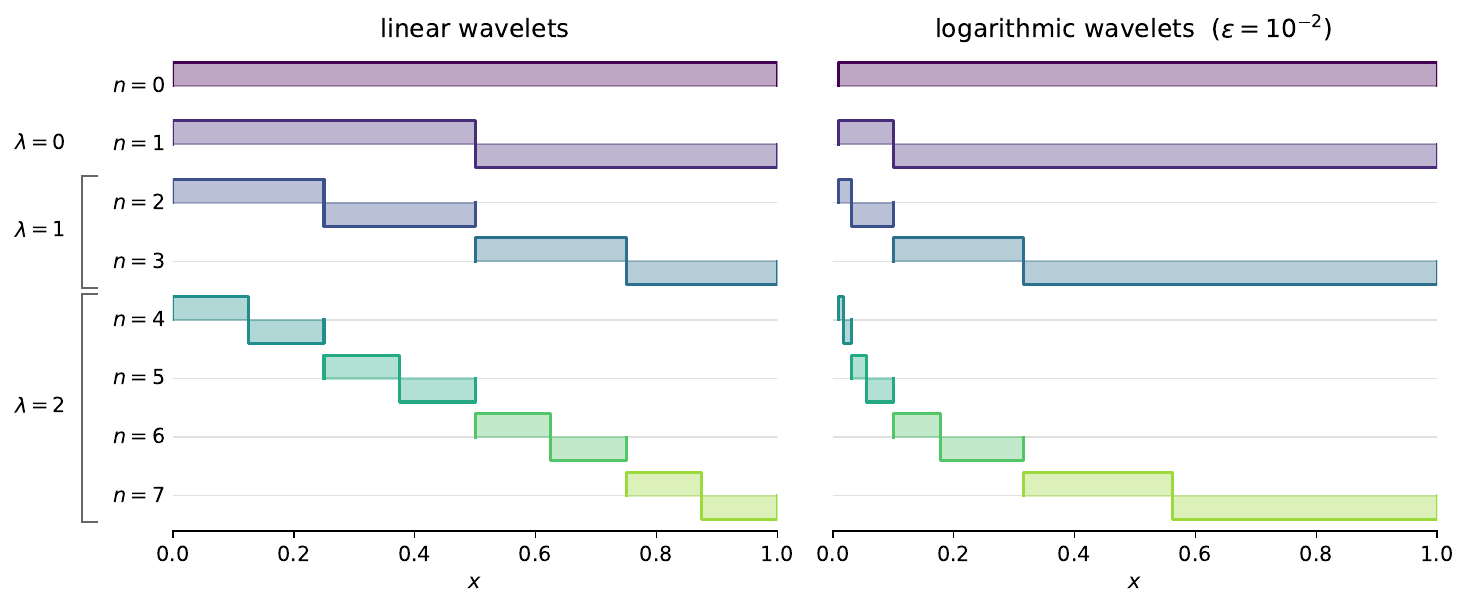}
\caption{The first eight radial basis functions of the linear (left) and logarithmic (right) spherical Haar wavelet bases, offset vertically by order $n$ and drawn on a common linear $x$ axis. Each wavelet is scaled to a common height so that its shape is visible; the true normalization factors $A_n$, $B_n$ are not shown. For the logarithmic basis we take $\epsilon = 10^{-2}$ for illustration. Both bases halve their supports generation by generation -- in $x$ for the linear basis, and in $\log x$ for the logarithmic basis -- so that the logarithmic wavelets concentrate toward small $x$, devoting equal resolution to every decade.}
\label{fig:wavelet_shapes}
\end{figure}

We keep the spherical Haar wavelet construction of \cref{sec:vsmethod} and change only the placement of the support boundaries. Recall that $h_0(x) = A_0$ is constant on the domain, while for $n \geq 1$ each wavelet is a pair of concatenated top-hats,
\begin{equation}
\label{eq:haar_def_log}
h_n(x) = \begin{cases}
+ A_n & x_{1,n} \leq x < x_{2,n} \,, \\
- B_n & x_{2,n} \leq x < x_{3,n} \,, \\
0 & \text{otherwise} \,,
\end{cases}
\end{equation}
with $n = 2^\lambda + \mu$ indexed by a generation $\lambda = 0, 1, \dots$ and position $\mu = 0, 1, \dots, 2^\lambda - 1$. In the \emph{logarithmic} basis, the boundaries of the $n \geq 1$ wavelets are uniformly spaced in $\log x$ between an IR cutoff $x = \epsilon$ and $x = 1$,
\begin{equation}
\label{eq:log_bounds}
\log x_{i,n} = -L + 2^{-\lambda} \Bigl(\mu + \frac{i-1}{2}\Bigr) L \,, \qquad i = 1, 2, 3 \,,
\end{equation}
where $L = \log(1/\epsilon)$ is the total logarithmic depth of the domain. The supports are halved generation by generation exactly as in the linear case, now in the variable $\log x$ on $[-L, 0]$ instead of $x$ on $[0,1]$. The orthonormality condition then gives
\begin{equation}
\label{eq:log_norm}
A_0 = \sqrt{\frac{3}{1 - \epsilon^3}} \,, \qquad
A_{n \geq 1} = \sqrt{\frac{3}{x_{1,n}^3} \, \frac{\rho_\lambda^3}{(\rho_\lambda^3 - 1)(\rho_\lambda^3 + 1)}} \,, \qquad
B_{n \geq 1} = \rho_\lambda^{-3} A_n \,,
\end{equation}
where $\rho_\lambda \equiv x_{2,n}/x_{1,n} = x_{3,n}/x_{2,n}$ (so that $\log \rho_\lambda = L/2^{\lambda+1}$) is the common ratio of the boundaries at generation $\lambda$. The first eight basis functions of the linear and logarithmic bases are compared in \cref{fig:wavelet_shapes}. When the basis is used for the material form factor, we set $\epsilon = q_\text{min}/q_\text{max}$. The VDF, by contrast, has no scale hierarchy, and we retain the linear basis on the velocity side throughout.

\subsection{Linear vs.\ logarithmic wavelets: convergence}
\label{sec:conv}

We now put the construction in the previous subsection to a quantitative test, comparing the convergence of the linear and logarithmic wavelet bases as the number of radial functions is increased. We consider realistic DM--phonon scattering calculations for both heavy and light mediators. The two cases probe opposite ends of the momentum range: the heavy mediator weights high $q$, while the light mediator concentrates the rate near $q_\text{min}$. Together they test whether the same basis can serve both mediator scenarios.

We compute the scattering rate in GaAs for the hadrophilic scalar mediator model, projecting the material form factor in each basis with up to $N_q=2048$ radial functions. We adopt a dedicated projection domain $\bigl[q_\text{min}, q_\text{max}\bigr]$ with $q_\text{max} = \min\bigl\{ 2m_\chi v_\text{max}\,, \, q_\text{cut} \bigr\}$ for each DM mass, as opposed to a common $q_\text{max}$ for all masses, in order to study the convergence without degradation of the effective resolution (which will be discussed in the next subsection). For the logarithmic basis, the IR cutoff $\epsilon = q_\text{min}/q_\text{max}$ then ranges from ${\cal O}(10^{-3})$ at $m_\chi = 0.1$~MeV to ${\cal O}(10^{-6})$ at $100$~MeV, assuming $\omega_\text{min} = 1$~meV. All other numerical parameters are identical for the two bases: $N_v = 128$ linear wavelets on the velocity side, $\ell_\text{max} = 5$, and energy bin width $\Delta\omega = 1$~meV. When computing the projection coefficients, we use grids of $128\times 180\times 180$ for the VDF and $2048\times 25\times 25$ for the material form factor in spherical coordinates, linearly (logarithmically) spaced in the radial direction for the linear (logarithmic) basis (note that at least 2048 grid points are necessary to compute $N_q=2048$ projection coefficients).

\begin{figure}[t]
\centering
\includegraphics[width=0.48\textwidth]{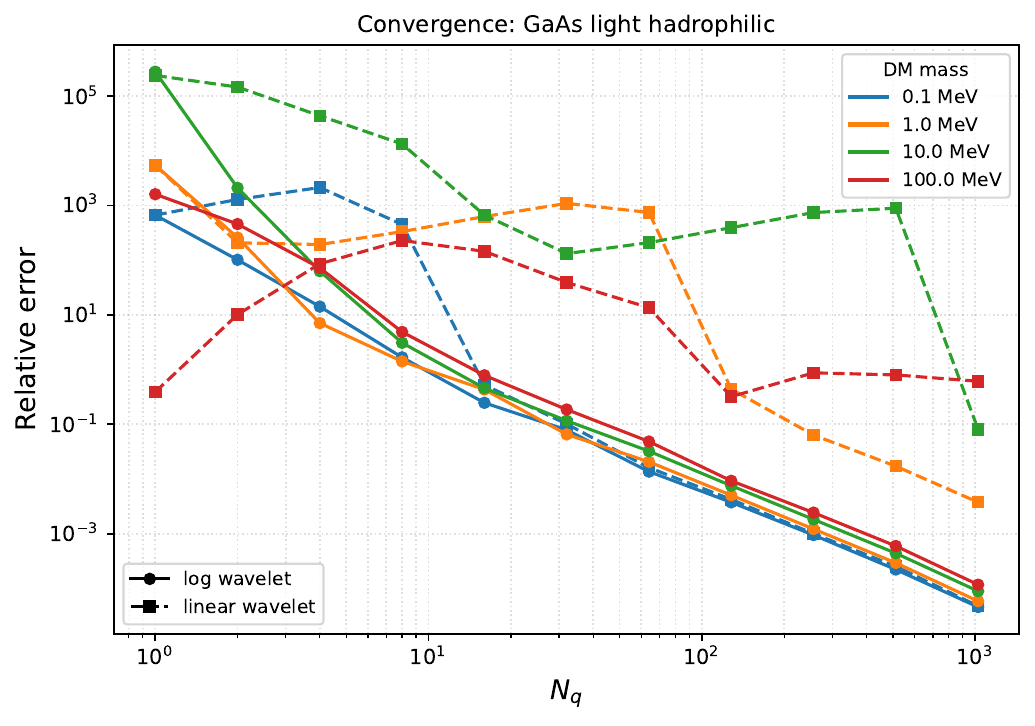}
\includegraphics[width=0.48\textwidth]{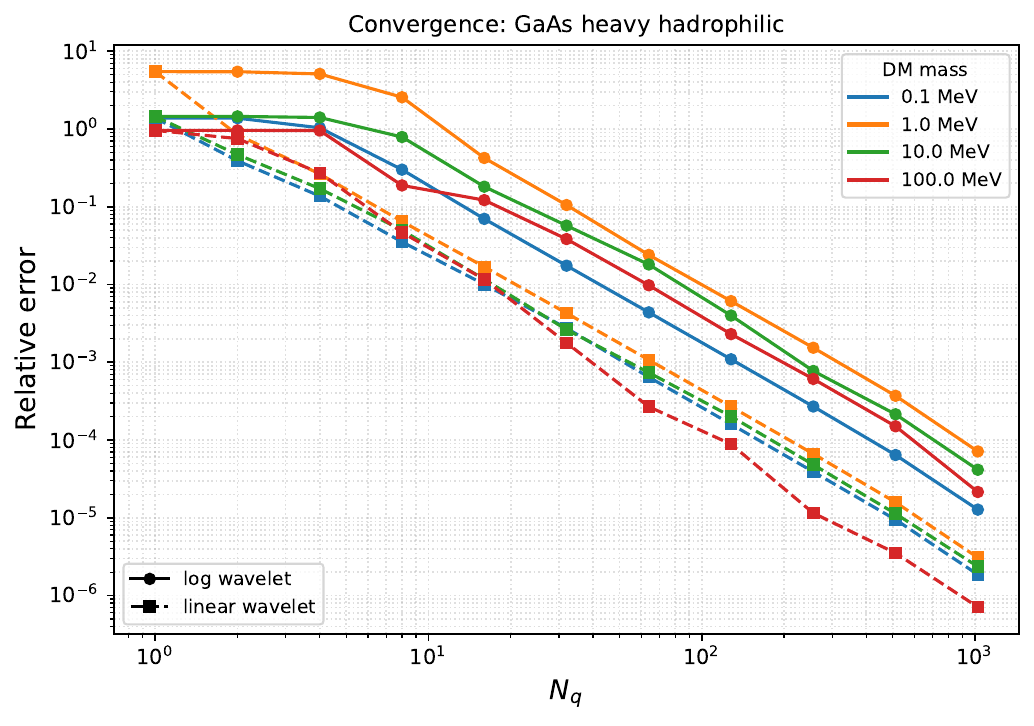}
\caption{Convergence of the rate with the number of radial basis functions $N_q$, for the linear (dashed) and logarithmic (solid) wavelet bases, in the light (left) and heavy (right) mediator cases (GaAs target, hadrophilic interaction). The relative error is defined with respect to the $N_q = 2048$ result within the same basis.}
\label{fig:wavelet_comparison}
\end{figure}

We define the relative error of the rate at a given $N_q$ with respect to the $N_q = 2048$ result within the same basis, and plot it as a function of $N_q$ for both light and heavy mediator cases and four DM masses spanning $0.1$--$100$~MeV in \cref{fig:wavelet_comparison}. Since truncating at $N_q = 2^\lambda$ retains all wavelet generations up to $\lambda - 1$, these curves can be read as a resolution budget: the drop in error from one $N_q$ to the next measures how much of the rate is carried by the newly added generation of coefficients.

For the light mediator, the contrast is dramatic. The logarithmic basis converges steadily and uniformly across all DM masses: the relative error from radial basis truncation is below $1\%$ at $N_q = 128$, and reaches $10^{-3}$ by $N_q=512$. The linear basis converges only for the lightest mass shown, $m_\chi = 0.1$~MeV, where the kinematically allowed momentum transfers span a relatively narrow range. As $m_\chi$ grows the range widens and the linear basis falls behind: at $m_\chi = 1$~MeV the relative errors are about two orders of magnitude worse than the logarithmic basis, and at $m_\chi = 100$~MeV it shows no sign of convergence at all, with relative errors close to or exceeding unity. This is the quantitative statement of the small-$q$ problem: uniform-width wavelets cannot resolve a $1/q^4$-weighted integrand near $q_\text{min}$ at any practical basis size.

For the heavy mediator, the linear basis fares better. The logarithmic basis nonetheless remains entirely usable: it requires only a factor of $\sim 4$ more basis functions to reach the same accuracy (e.g., $N_q = 256$ instead of $64$ to reach $10^{-3}$). This modest and affordable penalty is a key practical point: a single material form factor projection in the logarithmic basis serves both heavy and light mediator models, so the mediator mass need not be decided before the projection is computed. Faster convergence is not the only payoff of the logarithmic basis, however; the next subsection shows that it also removes the remaining obstacle to reusing one projection across a wide range of DM masses.

\subsection{Linear vs.\ logarithmic wavelets: one projection for all masses}
\label{sec:universal}

For a given DM mass, the range of momentum transfers $q$ contributing to the rate extends up to $\min\{2m_\chi v_\text{max}, q_\text{cut}\}$, which varies by four orders of magnitude over the phonon-accessible mass range. If we want to reuse a single material form factor projection across the full DM mass range, we have to set $q_\text{max}=q_\text{cut}$ to compute the projection coefficients $\langle n \ell m | f_{S,b}^2 \rangle$. This creates a numerical resolution issue: for a DM mass with $2m_\chi v_\text{max} < q_\text{cut}$, only the basis functions supported below $2m_\chi v_\text{max}$ carry information relevant to the rate. In the linear basis this is a fraction of the $N_q$ radial basis functions:
\begin{equation}
\label{eq:eff_fraction_linear}
\frac{N_q^\text{eff}}{N_q} \simeq \frac{2 m_\chi v_\text{max}}{q_\text{cut}} \equiv k(m_\chi) \qquad \text{(linear wavelets)}\,.
\end{equation}
When the projection is applied to sub-MeV masses, $k(m_\chi) \lesssim 10^{-2}$, so even $N_q = 512$ leaves essentially no usable resolution. Maintaining accuracy with the linear basis thus forces either an enormous $N_q$ or a series of projections tiling the DM mass range for each material, sacrificing the one-projection-for-all design principle of the vector space integration method.

The logarithmic basis changes the scaling. Its resolution is uniform per decade, so the effective fraction of basis functions covering the allowed region is set by the logarithmic rather than the linear measure:
\begin{equation}
\label{eq:eff_fraction}
\frac{N_q^\text{eff}}{N_q} \simeq \frac{\log k(m_\chi) - \log \epsilon}{\log 1 - \log \epsilon} = 1 - \frac{\log k(m_\chi)}{\log \epsilon} \qquad \text{(logarithmic wavelets)}\,,
\end{equation}
where $\epsilon = q_\text{min}/q_\text{cut}\sim 10^{-6}$ for $\omega_\text{min}\sim\mathcal{O}(\text{meV})$. For example, at $m_\chi = 0.1$~MeV, the effective fraction $N_q^\text{eff}/N_q$ is $\sim 50\%$ for the logarithmic basis, compared to $\sim 0.1\%$ for the linear basis.

\begin{figure}[t]
\centering
\includegraphics[width=0.48\textwidth]{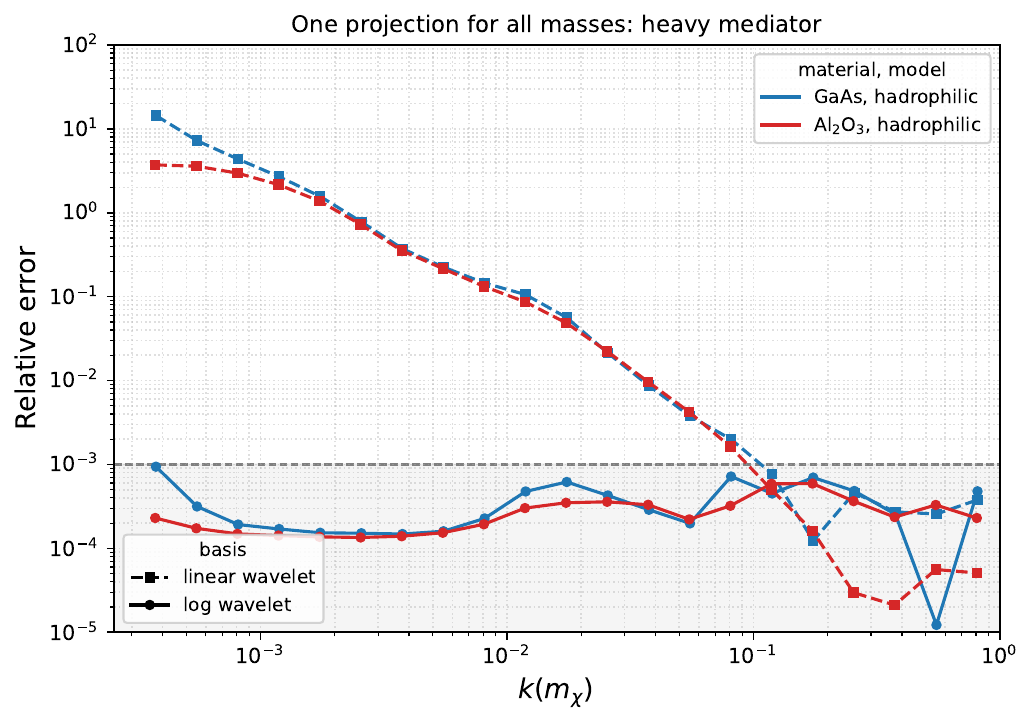}
\includegraphics[width=0.48\textwidth]{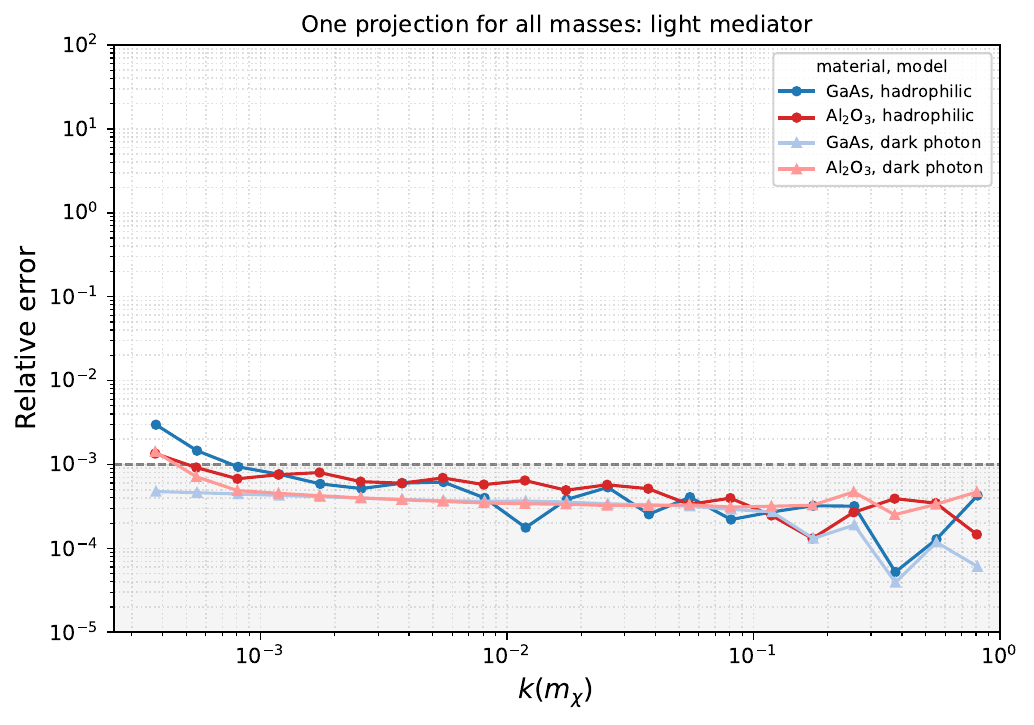}
\caption{Relative error of the rate computed from a single material form factor projection with $q_\text{max} = q_\text{cut}$ with respect to the reference rate computed from a dedicated projection with $q_\text{max} = 2m_\chi v_\text{max}$, as a function of $k(m_\chi) \equiv 2m_\chi v_\text{max}/q_\text{cut}$, for the heavy (left) and light (right) mediator models in GaAs and Al$_2$O$_3$. The two line styles in the heavy mediator case compare the linear (dashed) and logarithmic (solid) wavelet bases; for the light mediator only the logarithmic basis is shown. The shaded band indicates the $10^{-3}$ accuracy of the reference rate.}
\label{fig:qmax}
\end{figure}

\Cref{fig:qmax} quantifies this picture by measuring, across different DM masses, the accuracy of rates computed from a single material form factor projection with $q_\text{max}=q_\text{cut}$ and $N_q=512$ against rates computed from dedicated per-mass projections with $q_\text{max} = 2m_\chi v_\text{max}$ and $N_q = 128$ (linear) or $N_q = 512$ (logarithmic); the latter do not suffer from the effective resolution degradation, and serve as a reference. From \cref{fig:wavelet_comparison} we see that, with the exception of the linear basis for the light mediator where the calculation does not converge, these $N_q$ values chosen to compute the reference rates are sufficient to keep the relative error from radial basis truncation below $10^{-3}$, drawn as the shaded band in \cref{fig:qmax}. All other numerical parameters are the same as in \cref{sec:conv}.

In the linear basis (dashed curves in \cref{fig:qmax}), the error grows rapidly as $k(m_\chi)$ decreases, in line with the discussion around \cref{eq:eff_fraction_linear}: for the heavy mediator with $N_q = 512$, the $10^{-3}$ accuracy target is maintained only down to $k(m_\chi) \sim 0.1$, i.e., a single linear-basis projection covers roughly one decade of DM mass. In the logarithmic basis (solid curves), by contrast, the error remains at or below the $10^{-3}$ reference accuracy across the entire mass range. For the light mediator models, the linear basis does not yield a converged rate even with dedicated per-mass projections at any practical $N_q$ as we saw in \cref{fig:wavelet_comparison}, so we only show the results for the logarithmic basis. We see that the $10^{-3}$-level accuracy is maintained across almost the entire mass range, just as in the heavy mediator case.

The logarithmic wavelet basis thus resolves both obstacles identified at the beginning of this section. It converges for light mediators with $\mathcal{O}(100)$ radial basis functions, whereas the linear basis is estimated to need at least $\mathcal{O}(10^6)$, at the moderate cost of a factor $\sim 4$ in basis size for heavy mediators. It also makes the accuracy of a reused projection degrade only logarithmically in the DM mass. As a result, a single projection at $q_\text{max} = q_\text{cut}$ with $N_q = 512$ keeps the error from the truncation of the radial basis at the $\mathcal{O}(10^{-3})$ level over the entire phonon-accessible mass range. At this point, the angular grid size in the material form factor projection and the energy bin width (at the lowest DM masses) become the limiting factors for the accuracy of the rate; we quantify those in \cref{app:convergence}.

\section{\texttt{VectorPhonoDark}: DM--phonon scattering in practice}
\label{sec:vectorphonodark}

In this section, we present \texttt{VectorPhonoDark}~\cite{VectorPhonoDark}, the first Python package that provides the complete pipeline from DFT phonon data to rate predictions using the vector space integration method. We describe the workflow and design of the package in \cref{sec:workflow}, and the numerical choices behind each step in \cref{sec:implementation}. We then validate it against direct numerical integration in \cref{sec:performance}, on a daily modulation benchmark: the light dark photon mediator model with sapphire (Al$_2$O$_3$) as the target. The result agrees with \texttt{PhonoDark}~\cite{Trickle:2020oki,PhonoDark} at the $\lesssim 0.1\%$ level, while the computation time is reduced by orders of magnitude.

\subsection{Workflow}
\label{sec:workflow}

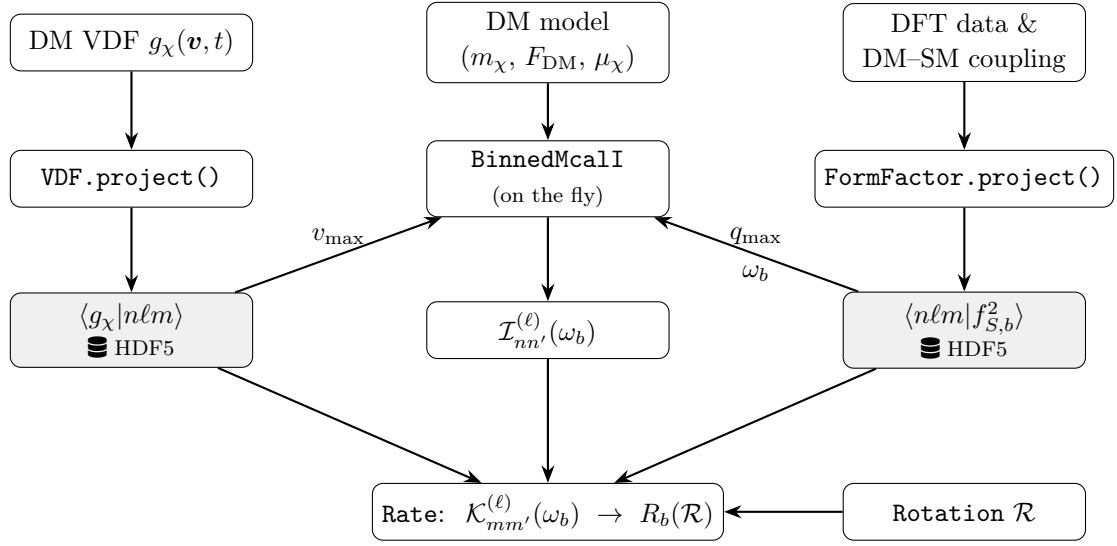
\begin{figure}[t]
\centering
\begin{tikzpicture}[
box/.style={draw, rounded corners, minimum width=3.2cm, minimum height=0.75cm, align=center, font=\small},
dbox/.style={draw, rounded corners, minimum width=3.2cm, minimum height=0.75cm, align=center, font=\small, fill=gray!12},
arrow/.style={->, >=Stealth, thick},
lbl/.style={font=\small}
]

\node[box] (vdf_func)  at (0,    0)   {DM VDF $g_\chi(\boldsymbol{v}, t)$};
\node[box] (vdf_proj)  at (0,   -1.8) {\texttt{VDF.project()}};
\node[dbox] (vdf_coef)  at (0,   -3.8) {$\langle g_\chi|n\ell m\rangle$\\[-1pt] {\scriptsize \faDatabase\ HDF5}};

\node[box] (dm_model) at (5.5,  0)   {DM model \\ $(m_\chi,\, F_\text{DM},\, \mu_\chi)$};
\node[box] (binned_km) at (5.5, -1.8) {\texttt{BinnedMcalI} \\ {\scriptsize (on the fly)}};
\node[box] (inn)       at (5.5, -3.8) {$\mathcal{I}_{nn'}^{(\ell)}(\omega_b)$};

\node[box] (phonopy)   at (11,   0)   {DFT data \&\\ DM--SM coupling};
\node[box] (ff_proj)   at (11,  -1.8) {\texttt{FormFactor.project()}};
\node[dbox] (ff_coef)  at (11,  -3.8) {$\langle n\ell m|f_{S,b}^2\rangle$\\[-1pt] {\scriptsize \faDatabase\ HDF5}};

\node[box] (rate)      at (5.5, -6.2) {\texttt{Rate}: $\ \mathcal{K}^{(\ell)}_{mm'}(\omega_b) \ \to\ R_b(\mathcal{R})$};
\node[box] (rotation)  at (11,  -6.2) {\texttt{Rotation} ${\cal R}$};

\draw[arrow] (vdf_func)  -- (vdf_proj);
\draw[arrow] (vdf_proj)  -- (vdf_coef);
\draw[arrow] (phonopy)   -- (ff_proj);
\draw[arrow] (ff_proj)   -- (ff_coef);
\draw[arrow] (dm_model)  -- (binned_km);
\draw[arrow] (binned_km) -- (inn);

\draw[arrow] (vdf_coef) -- node[above, lbl] {$v_\text{max}$} (binned_km);
\draw[arrow] (ff_coef) -- node[above, lbl] {$q_\text{max}$} node[below, lbl] {$\omega_b$} (binned_km);

\draw[arrow] (vdf_coef) -- (rate);
\draw[arrow] (inn)      -- (rate);
\draw[arrow] (ff_coef)  -- (rate);
\draw[arrow] (rotation) -- (rate);

\end{tikzpicture}
\caption{Workflow of the DM--phonon scattering rate calculation in \texttt{VectorPhonoDark}. The projections of the VDF (left) and of the binned material form factor (right) are computed once and stored as HDF5 files (shaded boxes) together with their full parameter metadata. The kinematic matrix (center) is analytic and computed on the fly for each DM mass and mediator. The rate (bottom) for any detector orientation or time of day is assembled by contracting the partial rate matrix with the Wigner $G$-matrices supplied by the \texttt{Rotation} helper class. When calculating the rate, the numerical parameters required for the kinematic matrix are imported automatically from the VDF and material form factor data.}
\label{fig:workflow}
\end{figure}

In \cref{eq:rate_master,eq:K_def}, the VDF, the material form factor, and the DM-model kinematics enter as independent factors, contracted only at the end. The package adopts this factorization directly as its architecture, with one class per ingredient:
\begin{itemize}
\item \textbf{VDF projection} (class \texttt{VDF}): the coefficients $\langle g_\chi | n \ell m \rangle$ of the DM velocity distribution.
\item \textbf{Material form factor projection} (class \texttt{FormFactor}): the coefficients $\langle n \ell m | f_{S,b}^2 \rangle$ of the binned material form factor, built from DFT phonon data and the DM--SM coupling.
\item \textbf{Kinematic scattering matrix} (class \texttt{BinnedMcalI}): the analytic matrices ${\cal I}_{nn'}^{(\ell)}(\omega_b)$ of \cref{eq:Ilnn} for a given DM model.
\item \textbf{Rate assembly} (class \texttt{Rate}): the contraction of the three ingredients into the partial rate matrix ${\cal K}^{(\ell)}_{mm'}(\omega_b)$ of \cref{eq:K_def}, and from it the binned rate $R_b({\cal R})$ for any rotation ${\cal R}$, represented by the helper class \texttt{Rotation}.
\end{itemize}
The real spherical harmonics used in both projections and the Wigner $G$-matrices used in the rate assembly are evaluated with \texttt{vsdm}~\cite{Lillard:2025aim}, so that our coefficient conventions agree with it. The main design decision left to the implementation is what to store and what to recompute. The two projections are the only steps that require numerical integration; they are comparatively expensive, and their outputs are compact coefficient arrays, so each is computed once and stored as HDF5 files. The kinematic scattering matrix is the opposite case: it is analytic and cheap to evaluate, but its $(\ell_\text{max} + 1) \times N_v \times N_q \times N_\text{bins}$ entries would take up considerable disk space, so it is rebuilt on the fly for each DM model. A scan therefore pays for the projections once and for one kinematic matrix per DM model, while detector orientations and times of day cost only the final contraction with the Wigner $G$-matrices $G^{(\ell)}({\cal R})$. The resulting workflow is illustrated in \cref{fig:workflow}.

The stored projections are designed as reusable data products rather than by-products of a single analysis. Each HDF5 file records the complete set of physical and numerical input parameters as metadata, so that any later calculation can verify compatibility before use. VDF projections computed with the \texttt{vsdm} package~\cite{Lillard:2025aim} follow the same coefficient conventions and can also be imported, provided these parameter attributes exist in the file. Since a projection is computed once per material and coupling type, and reused in every subsequent analysis, such files can accumulate into the shared library of material responses envisioned in Ref.~\cite{Lillard:2023qlx}; a parallel effort in the molecular channel~\cite{Blanco:2025sgv} follows the same pattern.

\subsection{Numerical implementation}
\label{sec:implementation}

\paragraph{Projections.}
The physical input to the VDF projection is the functional form of the VDF, together with its astrophysical parameters ($v_0$, $v_\text{E}$, $v_\text{esc}$ for the SHM). The material form factor projection instead starts from the DFT force constants and Born effective charges, which \texttt{phonopy}~\cite{phonopy-phono3py-JPCM,phonopy-phono3py-JPSJ} processes into the phonon frequencies $\omega_{\nu,\boldsymbol{k}}$ and polarization vectors $\boldsymbol{\epsilon}_{\nu,\boldsymbol{k},j}$ entering \cref{eq:fS2_phonon}, and from the DM--SM coupling, which fixes the effective charge $\boldsymbol{Y}_{\!j}$ of \cref{eq:Yj}. The routines that assemble the form factor are adapted from \texttt{PhonoDark}~\cite{Trickle:2020oki,PhonoDark}. The numerical inputs relevant for both projections include: the domain ($v_\text{max}$, or $(q_\text{min}, q_\text{max})$), the basis type and sizes ($N_v$ or $N_q$, and $\ell_\text{max}$),\footnote{Throughout this paper $N_v$ and $N_q$ denote the \emph{numbers} of radial basis functions, so that the radial index runs over $n = 0, 1, \dots, N-1$. The code instead takes the largest index as input, \texttt{n\_max} $= N - 1$; the values used here, $N_v = 128$ and $N_q = 512$, correspond to \texttt{nv\_max} $= 127$ and \texttt{nq\_max} $= 511$.} the energy bins in the material form factor case, and an integration grid of size $(N_r, N_\theta, N_\phi)$ in spherical coordinates.

The grid points are spaced uniformly in the radial coordinate for the linear basis, and uniformly in its logarithm for the logarithmic one. To keep the projection affordable, a single grid is generated and shared by all basis functions and, for the material form factor, by all energy bins: each grid point requires a \texttt{phonopy} evaluation of the phonon spectrum, which dominates the cost. The price of sharing is that higher-generation wavelets, whose supports are narrower, are sampled by fewer points. The package requires $N_r$ to be a power of two that is at least as large as $N_v$ (or $N_q$), so that every wavelet of the finest generation receives at least one grid point in each of its two intervals. We verify in \cref{app:convergence} that the recommended grids -- $(N_r, N_\theta, N_\phi) = (128, 180, 180)$ for the VDF and $(512, 25, 25)$ for the material form factor -- keep the grid-induced error below $\sim 0.6\%$ for all materials, couplings, and DM masses considered in this work.

Following \cref{sec:universal}, the recommended default for the material form factor is a single projection with $N_q = 512$ logarithmic wavelets on the domain $(q_\text{min}, q_\text{max}) = (\omega_\text{min}/v_\text{max},\, q_\text{cut})$, with $\omega_\text{min} = 1$~meV to accommodate the lowest energy thresholds of interest. It serves all DM masses and both heavy and light mediators, with the radial basis truncation error held at the $\lesssim 10^{-3}$ level over the entire mass range (\cref{sec:conv,sec:universal}). The coupling structure, by contrast, enters through $\boldsymbol{Y}_{\!j}$, so different couplings, such as in the hadrophilic scalar and dark photon mediator models, require separate projections of the same material. With the wavelet basis fixed in this way, the energy bin width $\Delta\omega$ is the only remaining parameter controlling the accuracy. The bin-center approximation error grows toward low DM masses, where the rate is increasingly dominated by the first few bins above threshold, and how quickly it grows depends on the material and the coupling. For $\Delta\omega = \omega_\text{min} = 1$~meV and the dark photon mediator, this error stays below $\sim 0.1\%$ in Al$_2$O$_3$ down to $m_\chi \sim 50$~keV. In GaAs, by contrast, it already reaches the percent level at $m_\chi \sim 100$~keV (see \cref{app:convergence} for details). At the lightest masses we therefore recommend a second projection with finer bins, e.g., $\Delta\omega = 0.1$~meV.

\paragraph{Rate calculation.}
The rate calculation takes as input the two stored projections, a DM model -- the DM mass, the mediator form factor, and the SM particle the DM couples to -- and a list of rotations. No numerical parameters are required at this stage: \texttt{Rate} reads $v_\text{max}$, $q_\text{max}$, the energy bins and the basis type from the VDF and material form factor files, so the kinematic scattering matrix is always built on the same basis as the projections it will be contracted with. These parameters are passed to \texttt{BinnedMcalI}, which evaluates ${\cal I}^{(\ell)}_{nn'}(\omega_b)$ at the bin centers from the closed-form solution of \cref{eq:Ilnn}, available from Ref.~\cite{Lillard:2023cyy} for mediator form factors of the monomial family\footnote{Compared to \cref{eq:FDM}, \cref{eq:fdm_qv} allows an extra monomial velocity dependence, which arises for some of the effective operators beyond the spin-independent interactions considered in this work. The closed-form solution in Ref.~\cite{Lillard:2023cyy} holds for arbitrary powers of both $q$ and $v$, so the implementation retains the general form. All benchmarks in this work use $b = 0$.}
\begin{equation}
\label{eq:fdm_qv}
F_\text{DM}^2(q, v) = \biggl(\frac{q}{q_\text{ref}}\biggr)^{\!a} \biggl(\frac{v}{c}\biggr)^{\!b} \,.
\end{equation}
For the spin-independent interactions with a heavy or light mediator considered in this paper, $a = 0$ or $-4$, and $b = 0$. Two implementations of ${\cal I}^{(\ell)}_{nn'}(\omega_b)$, both adapted from \texttt{vsdm}~\cite{Lillard:2025aim}, are provided: a compiled Cython~\cite{Behnel:2011cython} extension, and a Numba-compiled~\cite{Lam:2015numba} pure-Python fallback used automatically when the extension is unavailable. Detector orientations and times of day both enter as rotations of the crystal relative to the DM wind. Each rotation is represented by the class \texttt{Rotation}, built from an axis and an angle with \texttt{Rotation.from\_axis\_angle}, and acts on the crystal as an active rotation. It supplies the real Wigner $G$-matrices $G^{(\ell)}({\cal R})$ of \cref{eq:G_def}, which \texttt{Rate.binned\_rate} contracts with the partial rate matrix through \cref{eq:rate_master} to obtain the binned rate.

\subsection{Benchmark: daily modulation in Al$_2$O$_3$}
\label{sec:performance}

To validate the package and quantify its performance, we compute the daily modulation of the DM--phonon scattering rate in sapphire (Al$_2$O$_3$) for the light dark photon mediator model, and compare against direct numerical integration with \texttt{PhonoDark}~\cite{Trickle:2020oki,PhonoDark}. In the dark photon mediator model, the DM couples to electric charge, which enters the phonon rate through the Born effective charges of \cref{eq:Yj}. The reference cross section is conventionally normalized to DM--electron scattering, so $\mu_\chi$ is the DM--electron reduced mass and $q_\text{ref} = \alpha m_e$; the light mediator gives $a = -4$, $b=0$ in \cref{eq:fdm_qv}. Rates binned in energy with $\Delta\omega=1$~meV down to $\omega_\text{min} = 1$~meV are computed for four DM masses, $m_\chi = 0.05$, $0.1$, $0.5$, $1$~MeV, at $24$ hourly time steps. The crystal orientation at time $t$ is a rotation by $2\pi t/(24\,\text{h})$ about the Earth's rotation axis. We adopt the same setup as in Refs.~\cite{Griffin:2018bjn,Coskuner:2021qxo}, where the Earth's rotation axis lies at an angle $\theta_\text{E} \simeq 42^\circ$ from the DM wind, and the crystal $\hat{\boldsymbol{z}}$ axis is aligned with the wind at $t = 0$. \texttt{PhonoDark} integrates on its default logarithmically spaced momentum mesh with $100 \times 25 \times 25$ points, rebuilt at every time step. \texttt{VectorPhonoDark} uses an SHM projection ($\ell_\text{max} = 5$, $N_v = 128$, linear basis, $(N_r, N_\theta, N_\phi) = (128, 180, 180)$ grid) and a \emph{single} material form factor projection for all four masses: $q_\text{min} = \omega_\text{min}/v_\text{max}$, $q_\text{max} = q_\text{cut} \simeq 474$~keV, $N_q = 512$ logarithmic wavelets, $\Delta\omega = 1$~meV, $(N_r, N_\theta, N_\phi) = (512, 25, 25)$ grid.

\begin{figure}[t]
\centering
\includegraphics[width=0.72\textwidth]{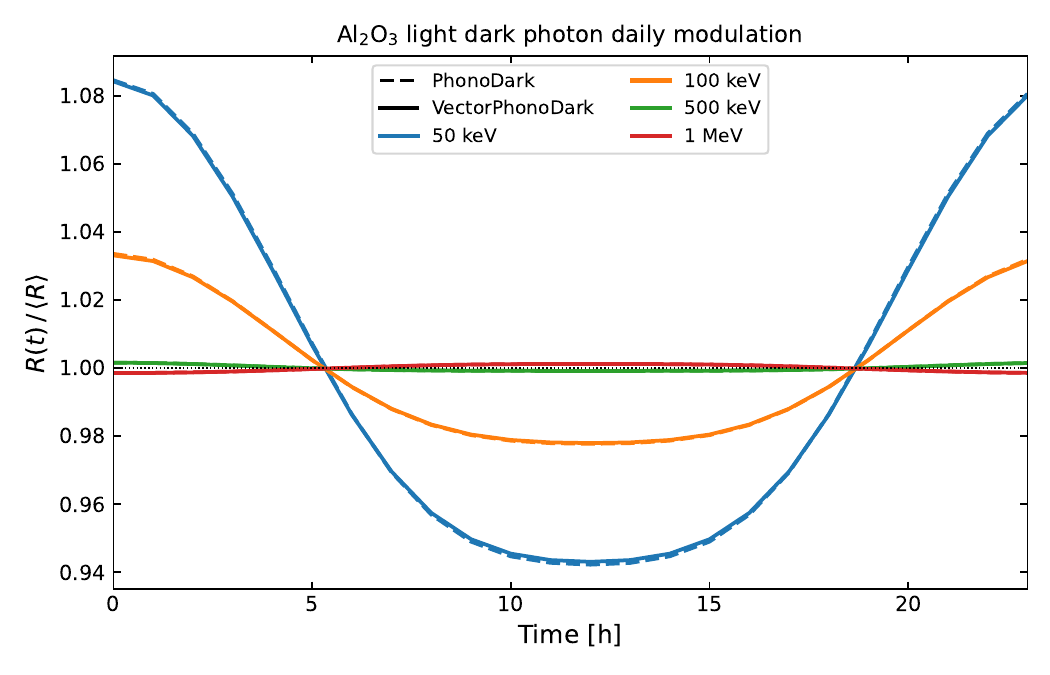}
\caption{Daily modulation of the DM--phonon scattering rate in Al$_2$O$_3$ for the light dark photon mediator model, normalized to the daily average, computed by direct numerical integration (\texttt{PhonoDark}, dashed) and by the vector space integration method (\texttt{VectorPhonoDark}, solid), for DM masses $m_\chi = 0.05$, $0.1$, $0.5$, $1$~MeV and an energy threshold of 20~meV. The \texttt{VectorPhonoDark} curves use a single material form factor projection, with $N_q = 512$ logarithmic wavelets, $q_\text{min} = \omega_\text{min}/v_\text{max}$ with $\omega_\text{min} = 1$~meV, $q_\text{max} = q_\text{cut}$, and $\Delta\omega = 1$~meV, for all four masses.}
\label{fig:modulation_al2o3_20meV}
\end{figure}

The modulation curves are shown in \cref{fig:modulation_al2o3_20meV}, and the corresponding comparison between \texttt{VectorPhonoDark} and \texttt{PhonoDark} in \cref{tab:modulation_20meV}. The daily-averaged rates agree to better than $0.05\%$ at every mass. The two codes also agree on the phase of the modulation, and the normalized curves $R(t)/\langle R \rangle$ agree pointwise to $0.1\%$ at $m_\chi = 0.05$~MeV and to better than $0.05\%$ for $m_\chi \geq 0.1$~MeV, well within the convergence errors detailed in \cref{app:convergence}. We have therefore validated both the vector space integration method in the phonon channel and its implementation in \texttt{VectorPhonoDark}.

\begin{table}[t]
\centering
\setlength{\tabcolsep}{9pt}
\renewcommand{\arraystretch}{1.2}
\begin{tabular}{cccc}
  \hline\hline
  \rule{0pt}{4.2ex}$m_\chi$
    & $\dfrac{\langle R\rangle^{\text{VPD}}}{\langle R\rangle^{\text{PD}}}$
    & $\dfrac{[R/\langle R\rangle]^{\text{VPD}}_{\text{min}}}{[R/\langle R\rangle]^{\text{PD}}_{\text{min}}}$
    & $\dfrac{[R/\langle R\rangle]^{\text{VPD}}_{\text{max}}}{[R/\langle R\rangle]^{\text{PD}}_{\text{max}}}$ \\[8pt]
  \hline
  $50\ \text{keV}$  & $1.0004$ & $1.0009$ & $0.9997$ \\
  $100\ \text{keV}$ & $0.9998$ & $1.0002$ & $0.9996$ \\
  $500\ \text{keV}$ & $0.9998$ & $1.0000$ & $1.0000$ \\
  $1\ \text{MeV}$   & $0.9998$ & $1.0000$ & $1.0000$ \\
  \hline\hline
\end{tabular}
\caption{Comparison of \texttt{VectorPhonoDark} (VPD) and \texttt{PhonoDark} (PD) for the Al$_2$O$_3$ daily modulation benchmark of \cref{fig:modulation_al2o3_20meV}. The second column gives the ratio of the daily-averaged rates; the remaining columns give the ratios of the minimum and the maximum of the modulation curve $R(t)/\langle R\rangle$.}
\label{tab:modulation_20meV}
\end{table}

\begin{table}[t]
\centering
\renewcommand{\arraystretch}{1.1}
\begin{tabular}{lr}
\hline\hline
VDF projection (once/halo) & $6$~s \\
Material form factor projection (once/material) & $211$~s \\
\hline
\multicolumn{2}{l}{Benchmark ($4$ masses $\times$ $24$ times)} \\
~~\texttt{VectorPhonoDark} & $9$~s \\
~~\texttt{PhonoDark} & $2.8$~h \\
\hline
Speedup & $\sim 1100 \times$ \\
\hline\hline
\end{tabular}
\caption{Wall-clock times for the Al$_2$O$_3$ daily modulation benchmark on an Apple M1 Pro, with both codes restricted to a single thread on a performance core.}
\label{tab:timing}
\end{table}

The computational cost is summarized in \cref{tab:timing}.\footnote{These results are obtained using \texttt{phonopy}~2.41.0, the last version of \texttt{phonopy} that the latest public version of \texttt{PhonoDark} supports. With a small compatibility patch to its \texttt{phonopy} interface, \texttt{PhonoDark} runs $\sim 1.4 \times$ faster with \texttt{phonopy} 4.4.0. Material form factor projection in \texttt{VectorPhonoDark} is also faster with \texttt{phonopy} 4.4.0, taking only $\sim 50$~s.} Once the one-time projections have been computed ($6$~s for the VDF, $211$~s for the material form factor), the entire benchmark of four DM masses and $24$ time steps completes in $9$~s. This cost is dominated by the four kinematic matrix evaluations, one per DM mass. Each additional time step is only a matrix multiplication, costing less than a millisecond. The same task by direct integration costs $1.4$--$2.0$~minutes \emph{per mass and time step}, or $2.8$~hours for the benchmark, so \texttt{VectorPhonoDark} achieves a speedup of $\sim 1100$. More importantly, the same projections carry over beyond this benchmark. Extending the scan to more masses costs $\sim 2$~s each, and switching from the light to the heavy mediator only requires new kinematic matrices. Furthermore, orientation studies with thousands of configurations are essentially free. With \texttt{PhonoDark}, by contrast, each such extension multiplies the cost: the full six-dimensional integration must be repeated for every new mass, mediator, and orientation. Together with the $\lesssim 0.1\%$ agreement established above, our \texttt{VectorPhonoDark} implementation makes systematic surveys over the DM parameter space, candidate materials, and detector orientations practical.

\section{Conclusions}
\label{sec:conclusion}

Single-phonon excitations in crystals extend the reach of direct detection down to DM masses of $\mathcal{O}(10)$~keV, and in anisotropic targets the rate also depends on the crystal orientation relative to the DM wind, modulating over a sidereal day. Optimizing the experimental sensitivity means computing the rate not once but many thousands of times -- across DM masses and mediator models, candidate materials, detector orientations, and times of day -- with every evaluation a six-dimensional integral over the DM's velocity and momentum transfer. The vector space integration method is designed for exactly this situation: the DM velocity distribution and the material form factor are each projected once onto a basis of wavelet-harmonic functions and stored in a shared library; the kinematics linking them reduces to an analytic matrix, and the rate for any DM model, orientation or time then follows from fast matrix algebra.

Applying the vector space integration method to the phonon channel, however, requires a new ingredient. The obstacle is a wide span of momentum scales peculiar to phonons, from $q_\text{min} \simeq 0.4$~eV determined by the energy threshold, to $q_\text{max} = q_\text{cut} \simeq 500$~keV set by the Debye--Waller cutoff of the phonon response. Haar wavelets whose supports are spaced uniformly in $q$, as used in existing implementations, resolve structure at a single scale. For light mediators, where the $1/q^4$ enhancement from the mediator propagator concentrates the rate just above $q_\text{min}$, convergence would demand an impractically large number of radial basis functions. Also, a single projection extending to $q_\text{cut}$ retains useful resolution only over the momentum transfers a given DM mass can actually reach, i.e., up to $2m_\chi v_\text{max}$, so its accuracy degrades for light DM, forcing a separate projection for every decade in mass and sacrificing the reusability that motivates the method.

We have shown that spacing the wavelet supports uniformly in $\log q$ instead overcomes the obstacle, because the resulting basis devotes equal resolution to every decade of momentum transfer. For light mediators it converges to the percent level with $\mathcal{O}(100)$ radial basis functions, and for heavy mediators it costs only a factor of $\sim 4$ more than the linear basis, so a single basis serves both mediator models. The accuracy of a reused projection, meanwhile, now degrades only logarithmically rather than linearly as the DM mass drops: one projection with $512$ logarithmic wavelets achieves $\lesssim 10^{-3}$ radial basis truncation error over the entire phonon-accessible mass range.

We have presented a new package \texttt{VectorPhonoDark}~\cite{VectorPhonoDark}, the first implementation of the vector space integration method for DM--phonon scattering. It covers the full pipeline: phonon frequencies and polarization vectors computed from \texttt{phonopy}~\cite{phonopy-phono3py-JPCM,phonopy-phono3py-JPSJ}, projection of the velocity distribution and of the energy-binned material form factor, on-the-fly construction of the analytic kinematic scattering matrix, and rate assembly for arbitrary detector orientations. Validating it against direct numerical integration with \texttt{PhonoDark}~\cite{Trickle:2020oki,PhonoDark}, on the daily modulation in Al$_2$O$_3$ for the light dark photon mediator model, we find agreement at the $\lesssim 0.1\%$ level, with the daily-averaged rates matching to better than $0.05\%$ at every mass. The benchmark of four DM masses and $24$ time steps takes $9$~s, against $2.8$~hours by direct integration, once the projections ($6$~s for the velocity distribution, $211$~s for the material form factor) are in hand. The projections are stored as HDF5 files with full parameter metadata, following the same coefficient conventions as \texttt{vsdm}~\cite{Lillard:2025aim}. Computed once per material and coupling type, they can accumulate into a shared library envisioned in Ref.~\cite{Lillard:2023qlx}. A parallel effort for molecular crystal targets~\cite{Blanco:2025sgv} suggests that such an ecosystem -- material-specific codes feeding a common vector space integration framework -- is taking shape across detection channels.

With the cost of each additional configuration reduced to matrix multiplications, analyses in the phonon channel that were previously time-consuming become routine: scans over halo model uncertainties~\cite{Li:2026xgj} and DM substructure beyond the standard halo model, the optimization of detector orientations for daily modulation searches, and comprehensive surveys of candidate crystals for optimal directional sensitivity. Extensions to general effective operators~\cite{Trickle:2020oki} and to multi-phonon responses~\cite{Campbell-Deem:2019hdx,Kahn:2020fef,Campbell-Deem:2022fqm,Lin:2023slv,Stratman:2024sng,Lin:2026qxr} are also natural next steps. More broadly, the logarithmic basis is not specific to phonons: any detection channel with a wide span of momentum scales can benefit from it, so the techniques developed here will have a lasting impact on the efficiency of DM direct detection calculations across the board.

\acknowledgments
We thank Ben Lillard for helpful discussions and feedback on the manuscript. X.-X.L.\ and Z.Z.\ are supported by the U.S. National Science Foundation under grant PHY-2412880. \texttt{VectorPhonoDark} was developed in collaboration with Claude Fable 5, Opus 4.6/4.7/4.8/5, Sonnet 4.6/5, GPT 5.4/5.5/5.6 Sol, and Gemini 3.1 Pro: after the initial code was written by X.-X.L., these AI tools were used to help with code refactoring, debugging, and algorithmic optimization, as well as providing useful feedback on earlier drafts of the manuscript. We also acknowledge the use of Get Physics Done~\cite{physical_superintelligence_2026_gpd} developed by Physical Superintelligence PBC (PSI), which provided useful feedback on the manuscript. This work was performed in part at the Aspen Center for Physics, which is supported by National Science Foundation grant PHY-1607611 and a grant from the Simons Foundation (1161654, Troyer).

\appendix
\crefalias{section}{appendix}

\section{Convergence}
\label{app:convergence}
\begin{figure}[t]
\centering
\includegraphics[width=0.45\textwidth]{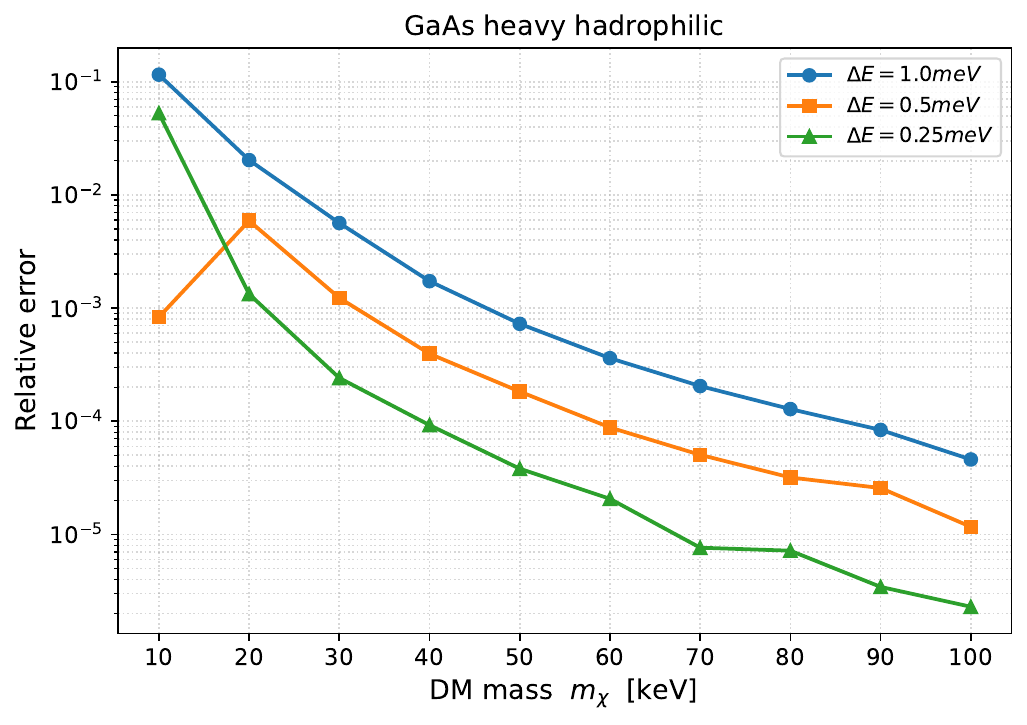}
\includegraphics[width=0.45\textwidth]{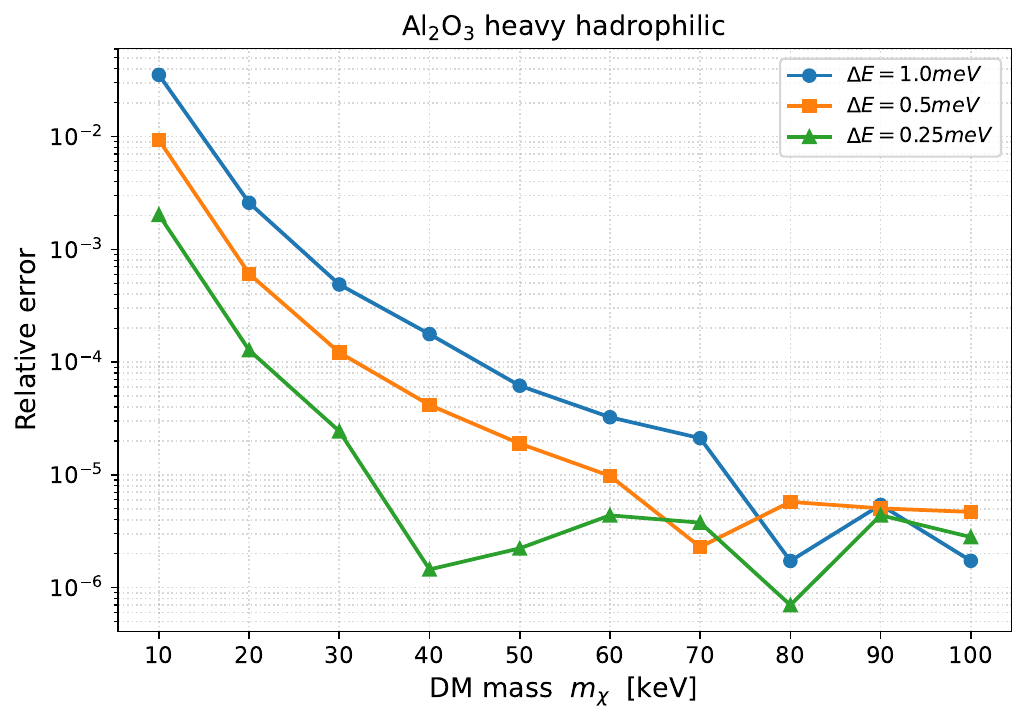}
\\
\includegraphics[width=0.45\textwidth]{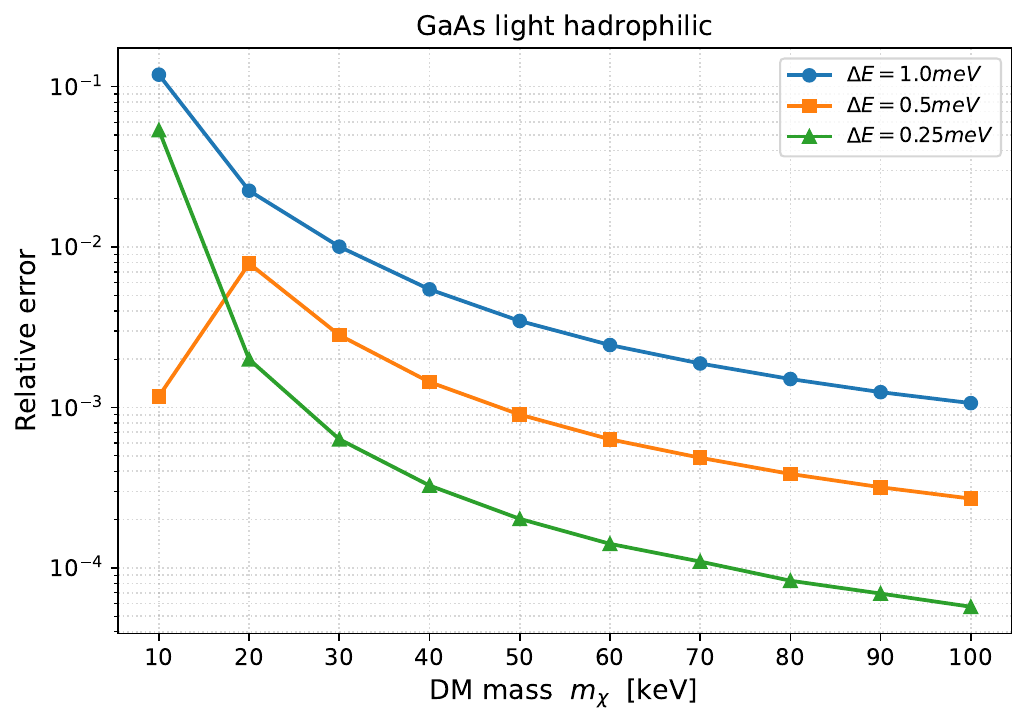}
\includegraphics[width=0.45\textwidth]{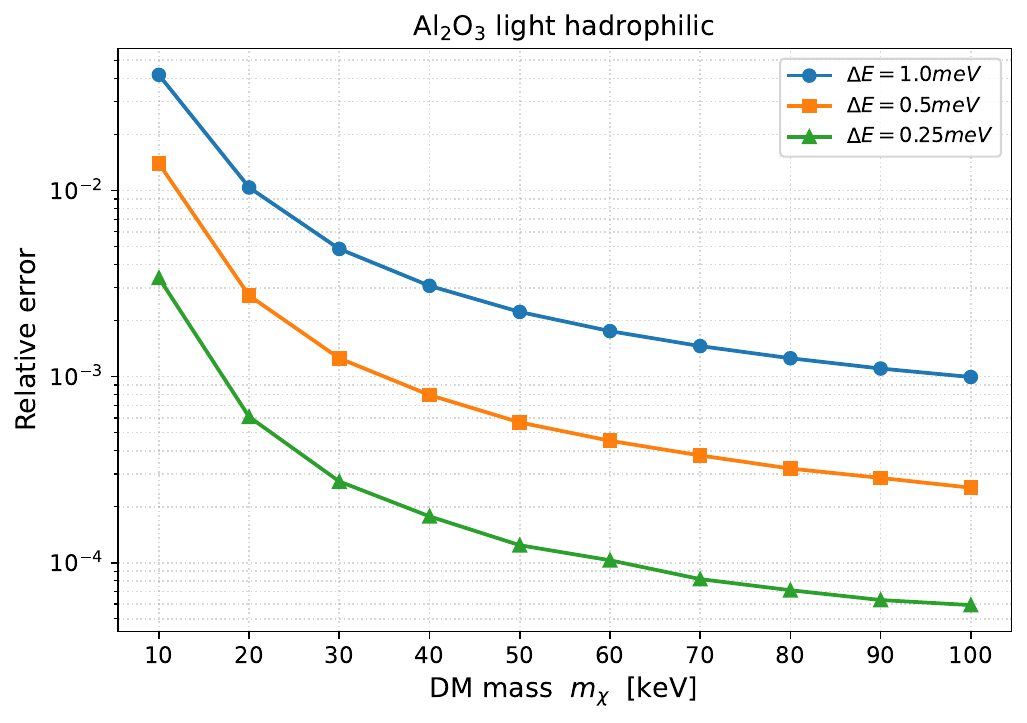}
\\
\includegraphics[width=0.45\textwidth]{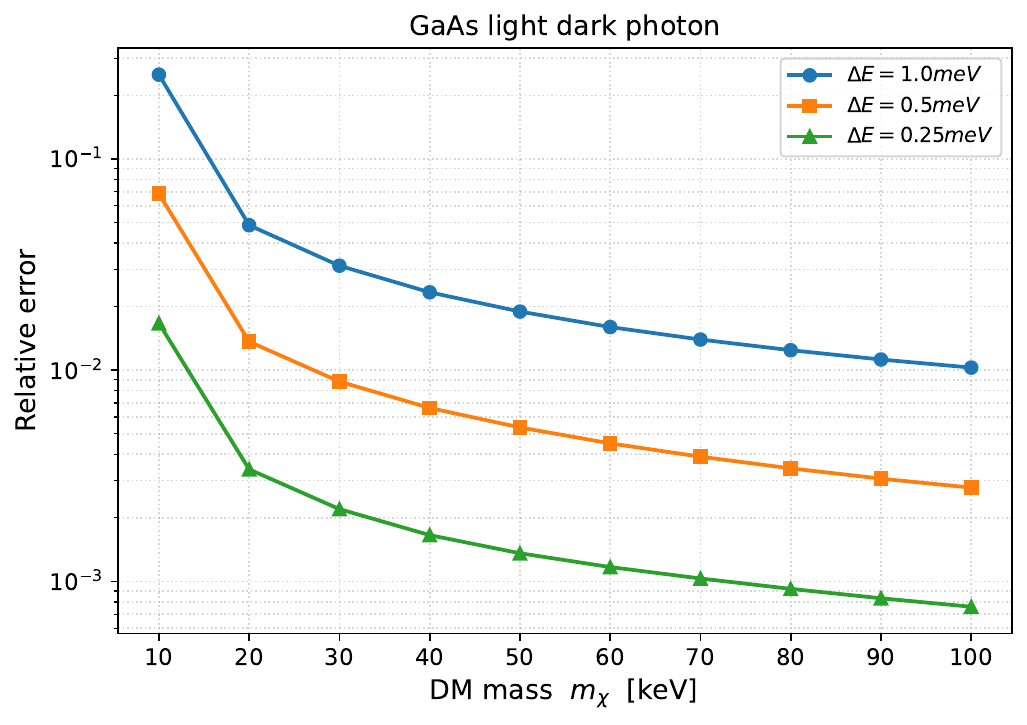}
\includegraphics[width=0.45\textwidth]{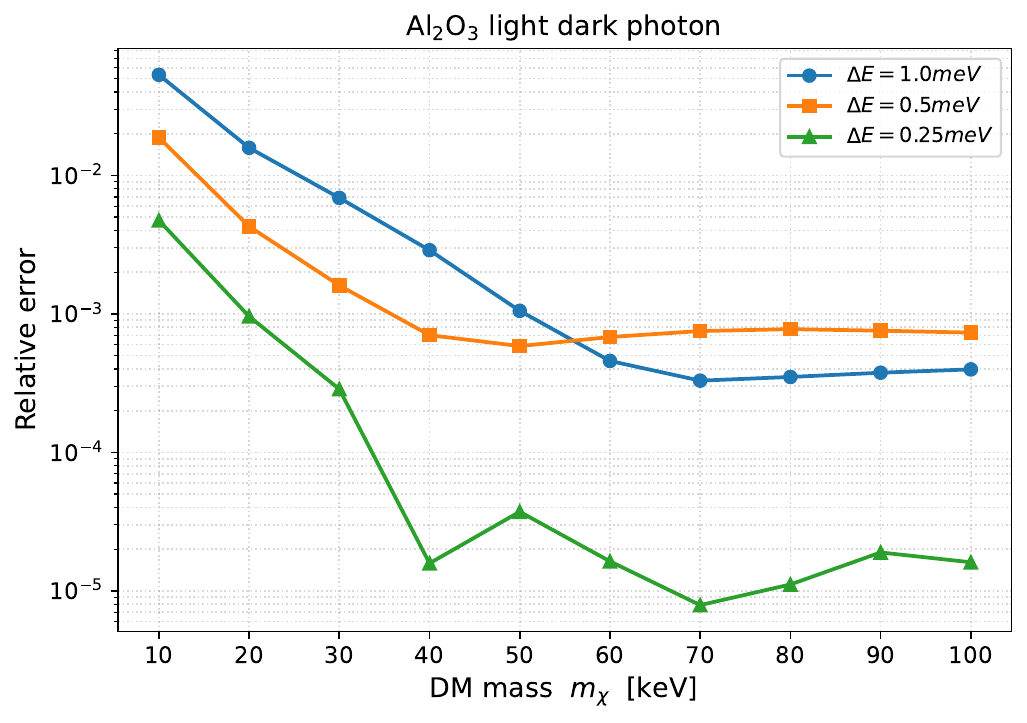}
\caption{Relative error in the rate for different energy bin widths, with respect to the reference $\Delta\omega = 0.1$~meV, for three mediator models (rows: heavy hadrophilic, light hadrophilic, light dark photon) and two materials (columns: GaAs, Al$_2$O$_3$). Unlike the rest of this appendix, this study uses $\ell_\text{max} = 5$; as \cref{tab:conv2_lmax} shows, the $\ell$ truncation affects the rate at or below the $0.01\%$ level, and is thus subdominant to the bin-width effects probed here.}
\label{fig:convergence_binwidth}
\end{figure}

We present in this appendix the results of the convergence study for the DM--phonon scattering rate calculation using \texttt{VectorPhonoDark}. The results are summarized in \cref{tab:conv2_lmax,tab:conv2_nvmax,tab:conv2_vdf_nr,tab:conv2_vdf_nang,tab:conv2_nqmax,tab:conv2_ff_nr_finest,tab:conv2_ff_nang} for the convergence in $\ell_\text{max}$ and in the projection parameters of the VDF and of the material form factor, and in \cref{fig:convergence_binwidth} for the convergence in the energy bin width $\Delta\omega$. Unless otherwise stated in the captions, the values for each case are computed with $\ell_\text{max}=8$, $N_v=128$, $N_q=512$, a $(N_r, N_\Omega) = (128, 180)$ grid for the VDF projection, a $(512, 25)$ grid for the material form factor projection, $\omega_\text{min} = 1$~meV, and $\Delta\omega=1$~meV. The two angular grid sizes are taken equal throughout this appendix, $N_\Omega \equiv N_\theta = N_\phi$. We choose these as the reference parameters because they sit comfortably in the converged regime for every quantity reported below. The convergence is studied for two materials (GaAs and Al$_2$O$_3$), three mediator models (heavy hadrophilic, light hadrophilic, and light dark photon), and four DM masses ($0.1$, $1$, $10$, and $100$~MeV).

\paragraph{Convergence in $\ell_\text{max}$.}
\Cref{tab:conv2_lmax} shows the relative error of the rate for $\ell_\text{max} = 2, 5$ against the reference $\ell_\text{max} = 8$. The expansion in real spherical harmonics converges remarkably fast: for both the nearly isotropic GaAs and the anisotropic Al$_2$O$_3$, even $\ell_\text{max} = 2$ keeps the error below $1\%$, and $\ell_\text{max} = 5$ yields $\leq 0.01\%$ error in all cases. This is the reason we adopt the modest setting $\ell_\text{max} = 5$ in \cref{sec:performance}, which keeps the partial rate matrix and the rotation tensors small without any visible impact on accuracy.

\begin{table}[tbp]
  \centering
  \renewcommand{\arraystretch}{1.2}
  \begin{tabular}{llccccc}
    \hline\hline
    Material & Model & $\ell_\text{max}$ &
      $0.1$~MeV & $1$~MeV &
      $10$~MeV & $100$~MeV \\
        \hline
    \multirow{6}{*}{GaAs}
      & \multirow{2}{*}{heavy, hadr.}
           & 2 & 0.00\% & 0.91\% & 0.01\% & 0.01\% \\
      &    & 5 & 0.00\% & 0.01\% & 0.00\% & 0.00\% \\
    \cline{2-7}
      & \multirow{2}{*}{light, hadr.}
           & 2 & 0.06\% & 0.04\% & 0.03\% & 0.04\% \\
      &    & 5 & 0.00\% & 0.01\% & 0.01\% & 0.01\% \\
    \cline{2-7}
      & \multirow{2}{*}{light, DP}
           & 2 & 0.03\% & 0.02\% & 0.01\% & 0.01\% \\
      &    & 5 & 0.00\% & 0.00\% & 0.00\% & 0.01\% \\
    \hline
    \multirow{6}{*}{Al$_2$O$_3$}
      & \multirow{2}{*}{heavy, hadr.}
           & 2 & 0.00\% & 0.12\% & 0.00\% & 0.01\% \\
      &    & 5 & 0.00\% & 0.00\% & 0.00\% & 0.00\% \\
    \cline{2-7}
      & \multirow{2}{*}{light, hadr.}
           & 2 & 0.01\% & 0.02\% & 0.02\% & 0.03\% \\
      &    & 5 & 0.00\% & 0.01\% & 0.01\% & 0.01\% \\
    \cline{2-7}
      & \multirow{2}{*}{light, DP}
           & 2 & 0.00\% & 0.02\% & 0.01\% & 0.02\% \\
      &    & 5 & 0.01\% & 0.00\% & 0.00\% & 0.00\% \\
    \hline\hline
  \end{tabular}
  \caption{Relative error in the rate vs.\ the reference ($\ell_\text{max}=8$) as a function of $\ell_\text{max}$, for all material/model combinations and four DM masses. ``hadr.''\ = hadrophilic; ``DP'' = dark photon. All other parameters are held at their reference values.}
  \label{tab:conv2_lmax}
\end{table}

\paragraph{Convergence in $N_v$ (radial VDF basis).}
\Cref{tab:conv2_nvmax} shows the convergence in the number of radial wavelets $N_v$ used to expand the VDF. The error decreases rapidly with $N_v$, reaching $\lesssim 0.02\%$ at $N_v = 128$ for all configurations. We therefore adopt $N_v = 128$ (\texttt{nv\_max} $= 127$) as the recommended value.

\begin{table}[htbp]
  \centering
  \renewcommand{\arraystretch}{1.2}
  \begin{tabular}{llccccc}
    \hline\hline
    Material & Model & $N_v$ &
      $0.1$~MeV & $1$~MeV &
      $10$~MeV & $100$~MeV \\
        \hline
    \multirow{6}{*}{GaAs}
      & \multirow{2}{*}{heavy, hadr.}
           & 64 & 0.06\% & 0.10\% & 0.07\% & 0.02\% \\
      &    & 128 & 0.01\% & 0.02\% & 0.01\% & 0.00\% \\
    \cline{2-7}
      & \multirow{2}{*}{light, hadr.}
           & 64 & 0.01\% & 0.02\% & 0.02\% & 0.02\% \\
      &    & 128 & 0.00\% & 0.00\% & 0.00\% & 0.00\% \\
    \cline{2-7}
      & \multirow{2}{*}{light, DP}
           & 64 & 0.02\% & 0.01\% & 0.02\% & 0.02\% \\
      &    & 128 & 0.00\% & 0.00\% & 0.00\% & 0.00\% \\
    \hline
    \multirow{6}{*}{Al$_2$O$_3$}
      & \multirow{2}{*}{heavy, hadr.}
           & 64 & 0.06\% & 0.09\% & 0.07\% & 0.02\% \\
      &    & 128 & 0.01\% & 0.02\% & 0.01\% & 0.00\% \\
    \cline{2-7}
      & \multirow{2}{*}{light, hadr.}
           & 64 & 0.02\% & 0.02\% & 0.02\% & 0.02\% \\
      &    & 128 & 0.00\% & 0.00\% & 0.00\% & 0.00\% \\
    \cline{2-7}
      & \multirow{2}{*}{light, DP}
           & 64 & 0.11\% & 0.00\% & 0.02\% & 0.02\% \\
      &    & 128 & 0.02\% & 0.00\% & 0.00\% & 0.00\% \\
    \hline\hline
  \end{tabular}
  \caption{Relative error in the rate vs.\ the reference ($N_v=256$ with $N_r^\text{VDF}=256$) as a function of $N_v$, for all material/model combinations and four DM masses. All other parameters are held at their reference values.}
  \label{tab:conv2_nvmax}
\end{table}

\paragraph{Convergence in the VDF projection grid.}
\Cref{tab:conv2_vdf_nr,tab:conv2_vdf_nang} show the dependence of the rate on the radial and angular sizes of the fixed grid used for the VDF projection. The radial direction is essentially saturated already at $N_r^\text{VDF} = 128$ ($\lesssim 0.01\%$ error), reflecting the smoothness of the SHM in $|\boldsymbol{v}|$. The angular direction requires more grid points: $N_\Omega^\text{VDF} = 90$ leaves residual $\lesssim 0.06\%$ errors, while $N_\Omega^\text{VDF} = 180$ brings them down to $\lesssim 0.01\%$. We recommend $(N_r, N_\Omega) = (128, 180)$ for the VDF projection.

\begin{table}[htbp]
  \centering
  \renewcommand{\arraystretch}{1.2}
  \begin{tabular}{llccccc}
    \hline\hline
    Material & Model & $N_r^\text{VDF}$ &
      $0.1$~MeV & $1$~MeV &
      $10$~MeV & $100$~MeV \\
        \hline
    \multirow{6}{*}{GaAs}
      & \multirow{2}{*}{heavy, hadr.}
           & 128 & 0.00\% & 0.01\% & 0.01\% & 0.00\% \\
      &    & 256 & 0.00\% & 0.01\% & 0.00\% & 0.00\% \\
    \cline{2-7}
      & \multirow{2}{*}{light, hadr.}
           & 128 & 0.00\% & 0.00\% & 0.01\% & 0.01\% \\
      &    & 256 & 0.00\% & 0.00\% & 0.00\% & 0.00\% \\
    \cline{2-7}
      & \multirow{2}{*}{light, DP}
           & 128 & 0.00\% & 0.00\% & 0.00\% & 0.01\% \\
      &    & 256 & 0.00\% & 0.00\% & 0.00\% & 0.00\% \\
    \hline
    \multirow{6}{*}{Al$_2$O$_3$}
      & \multirow{2}{*}{heavy, hadr.}
           & 128 & 0.00\% & 0.01\% & 0.01\% & 0.00\% \\
      &    & 256 & 0.00\% & 0.01\% & 0.00\% & 0.00\% \\
    \cline{2-7}
      & \multirow{2}{*}{light, hadr.}
           & 128 & 0.00\% & 0.01\% & 0.01\% & 0.01\% \\
      &    & 256 & 0.00\% & 0.00\% & 0.00\% & 0.00\% \\
    \cline{2-7}
      & \multirow{2}{*}{light, DP}
           & 128 & 0.01\% & 0.00\% & 0.00\% & 0.01\% \\
      &    & 256 & 0.01\% & 0.00\% & 0.00\% & 0.00\% \\
    \hline\hline
  \end{tabular}
  \caption{Relative error in the rate vs.\ the reference ($N_r^\text{VDF}=512$) as a function of the VDF radial grid size, for all material/model combinations and four DM masses. All other parameters are held at their reference values.}
  \label{tab:conv2_vdf_nr}
\end{table}

\begin{table}[htbp]
  \centering
  \renewcommand{\arraystretch}{1.2}
  \begin{tabular}{llccccc}
    \hline\hline
    Material & Model & $N_\Omega^\text{VDF}$ &
      $0.1$~MeV & $1$~MeV &
      $10$~MeV & $100$~MeV \\
        \hline
    \multirow{6}{*}{GaAs}
      & \multirow{2}{*}{heavy, hadr.}
           & 90 & 0.05\% & 0.05\% & 0.05\% & 0.02\% \\
      &    & 180 & 0.01\% & 0.01\% & 0.01\% & 0.00\% \\
    \cline{2-7}
      & \multirow{2}{*}{light, hadr.}
           & 90 & 0.02\% & 0.01\% & 0.01\% & 0.01\% \\
      &    & 180 & 0.00\% & 0.00\% & 0.00\% & 0.00\% \\
    \cline{2-7}
      & \multirow{2}{*}{light, DP}
           & 90 & 0.03\% & 0.02\% & 0.02\% & 0.01\% \\
      &    & 180 & 0.01\% & 0.01\% & 0.00\% & 0.00\% \\
    \hline
    \multirow{6}{*}{Al$_2$O$_3$}
      & \multirow{2}{*}{heavy, hadr.}
           & 90 & 0.05\% & 0.06\% & 0.05\% & 0.02\% \\
      &    & 180 & 0.01\% & 0.01\% & 0.01\% & 0.00\% \\
    \cline{2-7}
      & \multirow{2}{*}{light, hadr.}
           & 90 & 0.02\% & 0.01\% & 0.01\% & 0.01\% \\
      &    & 180 & 0.00\% & 0.00\% & 0.00\% & 0.00\% \\
    \cline{2-7}
      & \multirow{2}{*}{light, DP}
           & 90 & 0.06\% & 0.03\% & 0.02\% & 0.01\% \\
      &    & 180 & 0.01\% & 0.01\% & 0.00\% & 0.00\% \\
    \hline\hline
  \end{tabular}
  \caption{Relative error in the rate vs.\ the reference ($N_\Omega^\text{VDF}=360$) as a function of the VDF angular grid size, for all material/model combinations and four DM masses. All other parameters are held at their reference values.}
  \label{tab:conv2_vdf_nang}
\end{table}

\paragraph{Convergence in $N_q$ (radial material form factor basis).}
\Cref{tab:conv2_nqmax} shows the convergence in the number of radial wavelets $N_q$ used to expand the material form factor. The trend and the relative error are consistent with the partial results we have seen in \cref{sec:conv}, which are all below $0.1\%$ for $N_q \geq 512$. We therefore adopt $N_q = 512$ (\texttt{nq\_max} $= 511$) as the recommended value.

\begin{table}[htbp]
  \centering
  \renewcommand{\arraystretch}{1.2}
  \begin{tabular}{llccccc}
    \hline\hline
    Material & Model & $N_q$ &
      $0.1$~MeV & $1$~MeV &
      $10$~MeV & $100$~MeV \\
    \hline
    \multirow{6}{*}{GaAs}
      & \multirow{2}{*}{heavy, hadr.}
           & 512 & 0.01\% & 0.04\% & 0.02\% & 0.02\% \\
      &    & 1024 & 0.00\% & 0.01\% & 0.00\% & 0.00\% \\
    \cline{2-7}
      & \multirow{2}{*}{light, hadr.}
           & 512 & 0.02\% & 0.03\% & 0.04\% & 0.06\% \\
      &    & 1024 & 0.00\% & 0.01\% & 0.01\% & 0.01\% \\
    \cline{2-7}
      & \multirow{2}{*}{light, DP}
           & 512 & 0.02\% & 0.03\% & 0.05\% & 0.06\% \\
      &    & 1024 & 0.00\% & 0.01\% & 0.01\% & 0.01\% \\
    \hline
    \multirow{6}{*}{Al$_2$O$_3$}
      & \multirow{2}{*}{heavy, hadr.}
           & 512 & 0.01\% & 0.03\% & 0.02\% & 0.01\% \\
      &    & 1024 & 0.00\% & 0.01\% & 0.01\% & 0.00\% \\
    \cline{2-7}
      & \multirow{2}{*}{light, hadr.}
           & 512 & 0.02\% & 0.03\% & 0.04\% & 0.06\% \\
      &    & 1024 & 0.00\% & 0.01\% & 0.01\% & 0.01\% \\
    \cline{2-7}
      & \multirow{2}{*}{light, DP}
           & 512 & 0.02\% & 0.03\% & 0.05\% & 0.05\% \\
      &    & 1024 & 0.00\% & 0.01\% & 0.01\% & 0.01\% \\
    \hline\hline
  \end{tabular}
  \caption{Relative error in the rate vs.\ the reference ($N_q=2048$ with $N_r^\text{FF}=2048$) as a function of the number of radial wavelets $N_q$ used to expand the material form factor, for all material/model combinations and four DM masses. All other parameters are held at their reference values.}
  \label{tab:conv2_nqmax}
\end{table}

\paragraph{Convergence in the material form factor projection grid.}
\Cref{tab:conv2_ff_nr_finest,tab:conv2_ff_nang} show the convergence of the material form factor (FF) projection. \Cref{tab:conv2_ff_nr_finest} compares $N_r^\text{FF} = 512, 1024$ against the finest grid $N_r^\text{FF} = 2048$, demonstrating that $N_r^\text{FF} = 512$ already agrees with $N_r^\text{FF} = 2048$ to within $\sim 0.3\%$ across all masses and models, with most entries at or below the $0.1\%$ level. \Cref{tab:conv2_ff_nang} shows that $N_\Omega^\text{FF} = 25$ is sufficient: the residual errors are $\lesssim 0.6\%$ in the worst case (anisotropic Al$_2$O$_3$). We recommend $(N_r^\text{FF}, N_\Omega^\text{FF}) = (512, 25)$ for the material form factor projection.

\begin{table}[tbp]
  \centering
  \renewcommand{\arraystretch}{1.2}
  \begin{tabular}{llccccc}
    \hline\hline
    Material & Model & $N_r^\text{FF}$ &
      $0.1$~MeV & $1$~MeV &
      $10$~MeV & $100$~MeV \\
        \hline
    \multirow{6}{*}{GaAs}
      & \multirow{2}{*}{heavy, hadr.}
           & 512 & 0.01\% & 0.07\% & 0.10\% & 0.25\% \\
      &    & 1024 & 0.00\% & 0.01\% & 0.02\% & 0.03\% \\
    \cline{2-7}
      & \multirow{2}{*}{light, hadr.}
           & 512 & 0.01\% & 0.07\% & 0.04\% & 0.14\% \\
      &    & 1024 & 0.01\% & 0.03\% & 0.00\% & 0.01\% \\
    \cline{2-7}
      & \multirow{2}{*}{light, DP}
           & 512 & 0.02\% & 0.04\% & 0.03\% & 0.10\% \\
      &    & 1024 & 0.00\% & 0.01\% & 0.01\% & 0.01\% \\
    \hline
    \multirow{6}{*}{Al$_2$O$_3$}
      & \multirow{2}{*}{heavy, hadr.}
           & 512 & 0.01\% & 0.03\% & 0.15\% & 0.07\% \\
      &    & 1024 & 0.00\% & 0.01\% & 0.05\% & 0.08\% \\
    \cline{2-7}
      & \multirow{2}{*}{light, hadr.}
           & 512 & 0.01\% & 0.05\% & 0.01\% & 0.08\% \\
      &    & 1024 & 0.02\% & 0.03\% & 0.07\% & 0.03\% \\
    \cline{2-7}
      & \multirow{2}{*}{light, DP}
           & 512 & 0.02\% & 0.04\% & 0.06\% & 0.10\% \\
      &    & 1024 & 0.01\% & 0.01\% & 0.01\% & 0.05\% \\
    \hline\hline
  \end{tabular}
  \caption{Relative error in the rate vs.\ the reference ($N_r^\text{FF}=2048$) as a function of the material form factor radial grid size, for all material/model combinations and four DM masses. All other parameters are held at their reference values.}
  \label{tab:conv2_ff_nr_finest}
\end{table}

\begin{table}[htbp]
  \centering
  \renewcommand{\arraystretch}{1.2}
  \begin{tabular}{llccccc}
    \hline\hline
    Material & Model & $N_\Omega^\text{FF}$ &
      $0.1$~MeV & $1$~MeV &
      $10$~MeV & $100$~MeV \\
        \hline
    \multirow{6}{*}{GaAs}
      & \multirow{2}{*}{heavy, hadr.}
           & 15 & 0.24\% & 0.14\% & 0.80\% & 0.43\% \\
      &    & 25 & 0.07\% & 0.04\% & 0.09\% & 0.23\% \\
    \cline{2-7}
      & \multirow{2}{*}{light, hadr.}
           & 15 & 0.07\% & 0.07\% & 0.00\% & 0.13\% \\
      &    & 25 & 0.04\% & 0.03\% & 0.04\% & 0.00\% \\
    \cline{2-7}
      & \multirow{2}{*}{light, DP}
           & 15 & 0.06\% & 0.10\% & 0.01\% & 0.02\% \\
      &    & 25 & 0.02\% & 0.03\% & 0.04\% & 0.03\% \\
    \hline
    \multirow{6}{*}{Al$_2$O$_3$}
      & \multirow{2}{*}{heavy, hadr.}
           & 15 & 0.13\% & 0.06\% & 1.61\% & 1.58\% \\
      &    & 25 & 0.04\% & 0.04\% & 0.60\% & 0.46\% \\
    \cline{2-7}
      & \multirow{2}{*}{light, hadr.}
           & 15 & 0.02\% & 0.19\% & 0.07\% & 0.13\% \\
      &    & 25 & 0.02\% & 0.01\% & 0.03\% & 0.02\% \\
    \cline{2-7}
      & \multirow{2}{*}{light, DP}
           & 15 & 0.67\% & 0.00\% & 0.18\% & 0.54\% \\
      &    & 25 & 0.28\% & 0.02\% & 0.06\% & 0.05\% \\
    \hline\hline
  \end{tabular}
  \caption{Relative error in the rate vs.\ the reference ($N_\Omega^\text{FF}=50$) for the material form factor projection, as a function of the material form factor angular grid size, for all material/model combinations and four DM masses. All other parameters are held at their reference values.}
  \label{tab:conv2_ff_nang}
\end{table}

\paragraph{Convergence in the energy bin width.}
Finally, \cref{fig:convergence_binwidth} shows the convergence with respect to the energy bin width, comparing $\Delta\omega = 1$, $0.5$, $0.25$~meV against the reference $\Delta\omega = 0.1$~meV. The study is restricted to DM masses $m_\chi \leq 100$~keV, where the bin-width effect is most pronounced. The dominant trend is with the DM mass: the error grows rapidly as $m_\chi$ decreases. This is because the accessible range of phonon energies shrinks toward the threshold, so the rate comes to sit in the lowest energy bins, where the fractional bin width $\Delta\omega/\omega_b$ -- and hence the bin-center approximation error -- is largest. At fixed $m_\chi$ the effect also depends on the material and the coupling, and is largest for the dark photon mediator model. For GaAs with the dark photon mediator, the $\Delta\omega = 1$~meV error remains at the $\sim 1\%$ level out to $m_\chi \approx 100$~keV. For the hadrophilic mediator models and for Al$_2$O$_3$, it falls below $\sim 0.1\%$ already by $m_\chi \approx 50$--$100$~keV. As a practical guideline, $\Delta\omega = 1$~meV suffices for $m_\chi \gtrsim 100$~keV in all but the GaAs dark photon mediator case, while for lighter masses $\Delta\omega = 0.1$~meV is recommended.

\phantomsection
\addcontentsline{toc}{section}{References}

\bibliographystyle{JHEP}
\bibliography{ref}

\end{document}